\documentclass[journal,final,twocolumn,10pt,twoside]{IEEEtranTCOM}

\usepackage{stfloats}

\normalsize

\ifCLASSINFOpdf
 
\else
 
\fi

\usepackage{mathtools}
\usepackage{amsmath}
\usepackage{amssymb}
\usepackage{amsmath}
\usepackage{cite}
\usepackage{siunitx}
\usepackage{multirow}
\usepackage{multicol}

\usepackage{graphicx}
\usepackage{graphicx}
\date{}

\usepackage{calc}
\newcolumntype{M}[1]{>{\centering\arraybackslash}m{#1}}
\newcolumntype{N}{@{}m{0pt}@{}}

\usepackage{mdwmath}
\usepackage{blindtext}
\usepackage{eqparbox}
\usepackage{fixltx2e}
\usepackage{stfloats}
\DeclareMathOperator{\E}{\mathbb{E}}
\usepackage{slashbox}
\begin{document}
\title{MIMO Underwater Visible Light Communications:\\ Comprehensive Channel Study, Performance Analysis, and Multiple-Symbol Detection}
\author{Mohammad~Vahid~Jamali, Pooya Nabavi, and~Jawad~A.~Salehi,~\IEEEmembership{Fellow,~IEEE}\\
\thanks{Mohammad Vahid Jamali is with the Electrical Engineering and Computer Science Department, University of Michigan, Ann Arbor, MI, USA (e-mail: mvjamali@umich.edu).
Pooya Nabavi is with the Department of Electrical and Computer Engineering, Rice University, Houston, TX, USA (e-mail: pooya.nabavi@gmail.com).
 And Jawad A. Salehi is with the Optical Networks Research Laboratory (ONRL), Department of Electrical Engineering, Sharif University of Technology, Tehran, Iran (e-mail: jasalehi@sharif.edu).}

 \vspace{-0.8cm}
}

\maketitle
\begin{abstract}
\boldmath
This paper presents a comprehensive study of underwater visible light communications (UVLC), from channel characterization, performance analysis, and effective transmission and reception methods. To this end, we first simulate the fading-free impulse response (FFIR) of UVLC channels using Monte Carlo numerical procedure to take into account the absorption and scattering effects; and then to characterize turbulence effects, we multiply the aforementioned FFIR by a fading coefficient which for weak oceanic turbulence can be modeled as a lognormal random variable (RV). Based on this general channel model, we analytically study the bit error rate (BER) performance of UVLC systems with binary pulse position modulation (BPPM). 
In the next step, to mitigate turbulence effects, we employ multiple transmitters and/or receivers, i.e., we apply spatial diversity technique over UVLC links. Closed-form expressions for the system BER are provided, when an equal gain combiner (EGC) is employed at the receiver side, thanks to the Gauss-Hermite quadrature formula as well as approximation to the sum of lognormal RVs. We further apply saddle-point approximation, an accurate photon-counting method, to evaluate the system BER in the presence of shot noise. Both laser-based collimated and light emitting diode (LED)-based diffusive links are investigated. 
Additionally, in order to reduce the inter-symbol interference (ISI), introduced by the multiple-scattering effect of UVLC channels on the propagating photons, we also obtain the optimal multiple-symbol detection (MSD) algorithm, as well as the sub-optimal generalized MSD (GMSD) algorithm.
Our numerical analysis indicate good matches between the analytical and photon-counting results implying the negligibility of signal-dependent shot noise, and also between the analytical results and numerical simulations confirming the accuracy of our derived closed-form expressions for the system BER. Besides, our results show that spatial diversity significantly mitigates fading impairments while (G)MSD considerably alleviates ISI deterioration.
\vspace{0ex}
\end{abstract}
\section{Introduction}
\IEEEPARstart{D}{ue} to its unique advantages, underwater
visible light communications (UVLC) is receiving growing attention as a dominant scheme for high-throughput short-range underwater wireless communications. Compared to its well-investigated counterpart, namely acoustic communications, UVLC has many superiorities including higher bandwidth, lower time latency, and better security. Moreover, UVLC systems are relatively cost-effective and easy-to-install. Thanks to these peerless advantages, UVLC can be considered as an alternative to meet the requirements of high-speed and large-data underwater communications and to be applied in various underwater applications such as imaging, real-time video transmission, high-throughput sensor networks, etc. \cite{tang2014impulse,hanson2008high}. Despite all these advantages, several phenomena, namely absorption, scattering, and turbulence, adversely affect the photons' propagation under water. These factors cause loss, inter-symbol interference (ISI), and fading on the received optical signal, respectively, and limit the viable communication range of UVLC systems to typically shorter than $100$ \si{m}. This impediment hampers on the widespread usage of UVLC systems for longer ranges and necessitates intelligent system design and efficient transmission and reception methods.

Considerable research activities, both theoretically and experimentally, have been accomplished to characterize absorption and scattering effects of different water types \cite{mobley1994light,petzold1972volume}. Modeling of a UVLC channel and its performance evaluation using radiative transfer theory have been presented in \cite{jaruwatanadilok2008underwater}. Based on the experimental results reported in \cite{mobley1994light,petzold1972volume}, Tang \textit{et al.} \cite{tang2014impulse} used Monte Carlo (MC) approach to simulate the fading-free impulse response (FFIR) of UVLC channels with respect to absorption and scattering effects. They also fitted a double gamma function (DGF) to this impulse response and numerically evaluated the system bit error rate (BER) without considering turbulence effects. Also, the FFIR of multiple-input multiple-output (MIMO) UVLC systems has recently been simulated in \cite{zhang2016impulse} and a weighted Gamma function polynomial (WGFP) has been proposed to model the FFIR of MIMO-UVLC links with arbitrary number of light sources and detectors.
Moreover, in \cite{akhoundi2015cellular} a cellular code division multiple access (CDMA) UVLC network has been introduced based on assigning a unique optical orthogonal code (OOC) to each underwater mobile user. Meanwhile, potential applications and challenges of the aforementioned underwater cellular optical CDMA network, and the beneficial application of serial relaying on its users' performance have very recently been investigated in \cite{akhoundi2016cellular} and \cite{jamali2016relay}, respectively. In the
mean time, performance analysis of multi-hop UVLC systems with respect to all of the channel
degrading effects can be found in \cite{jamali2016performance}.
    
On the other hand, optical turbulence, which results as a consequence of random variations in the water refractive index, causes fluctuations and fading on the received optical signal. This phenomenon is called turbulence-induced fading and adversely affects the  performance of UVLC systems, especially for longer link ranges. Therefore, the precise estimation of the performance of UVLC systems requires accurate and detailed characterization of underwater optical turbulence. 
Recently, some useful results have been reported in the literature on characterizing underwater turbulence. An accurate power spectrum has been derived for fluctuations of turbulent seawater refractive index \cite{nikishov2000spectrum}.
 Based on this power spectrum and Rytov method, the scintillation index of optical plane and spherical waves propagating in underwater turbulent medium have been evaluated in \cite{korotkova2012light,ata2014scintillations}. In \cite{gerccekciouglu2014bit}, the on-axis scintillation index of a focused Gaussian beam has been formulated in weak oceanic turbulence and, by considering lognormal distribution for intensity fluctuations, the average BER is evaluated. 

This research, as a comprehensive work, aims to thoroughly investigate the UVLC channel for different channel conditions and system configurations, and then suggest effective solutions as intelligent transmission and reception methods to alleviate the channel impediments and extend the boundaries. Therefore, we first extensively study the UVLC channel for both collimated and diffusive links and investigate the channel temporal and spatial spread in various conditions. Then as a potent method for mitigating turbulence-induced fading, as a serious channel impairment, we apply MIMO transmission and analytically evaluate the performance of MIMO-UVLC systems with respect to all of the channel degrading effects. Finally, multiple-symbol detection (MSD) and generalized MSD (GMSD) algorithms, as effective detection methods for ISI channels, will be investigated. A significant advantage of binary pulse position modulation (BPPM) compared to on-off keying (OOK) modulation is that the detection process in BPPM does not require any channel state information (CSI). Therefore, all of the derived detection algorithms in this paper, even (G)MSD, are based on the absence of CSI.
        
 The remainder of this paper is organized as follows. In Section II, we review the UVLC channel modeling, including MC simulation method for collimated and diffusive links as well as underwater turbulence characteristics, followed by describing the BPPM-based MIMO-UVLC system with equal gain combiner (EGC). In Section III, we analytically calculate the BER expressions for both single-input single-output (SISO) and MIMO configurations. The BER closed-form solutions are also obtained in this section using Gauss-Hermite quadrature formula. In Section IV, we apply saddle-point approximation to evaluate the system BER in the presence of shot noise using photon-counting method. Section V derives the (G)MSD algorithm(s) for BPPM UVLC systems, Section VI presents the numerical results for various link configurations and system parameters, and Section VII concludes the paper.
\section{Channel and System Model}
\subsection{FFIR Simulation Using Monte Carlo Method}
 Through the propagation of optical beam under water, interactions between each photon and seawater particles causes  absorption and scattering phenomena. Absorption is an irreversible process where photons interact with water molecules and other particles and thermally lose their energy. In the scattering process, this interaction alters the propagation direction of each photon which, in addition to ISI, can also cause energy loss since fewer photons will be captured by the receiver aperture. Energy loss of non-scattered light due to absorption and scattering can be characterized by absorption coefficient $a(\lambda )$ and scattering coefficient $b(\lambda )$, respectively. And the total effect of absorption and scattering on energy loss can be described by extinction coefficient $c\left(\lambda \right)=a\left(\lambda \right)+b(\lambda )$. These coefficients can vary with the source wavelength $\lambda $ and water type \cite{tang2014impulse}. It has been shown in \cite{mobley1994light} and \cite{bohren2008absorption} that absorption and scattering have their lowest effects at the interval $400$ \si{nm} $<\lambda <530$ \si{nm}; thus, UVLC systems apply the blue/green region of the visible light spectrum to actualize underwater optical data communications.

In order to take into account the absorption and scattering effects of UVLC channels, we use MC numerical method to simulate the channel impulse response regardless of turbulence effects, in a similar approach to  \cite{tang2014impulse,cox2012simulation,cox2014simulating}. We name this impulse response as FFIR of the channel and denote it by $h_{0,ij}(t)$ for the channel between the $i$th transmitter and $j$th receiver.

The basic steps of FFIR simulation using MC method can be summarized as follows. First, we generate numerous photons at the transmitter, each with assigned initial attributes. Specific attributes of each photon include their position in Cartesian coordinates $(x,y,z)$, the transmission direction described by zenith angle $\theta$ and azimuth angle $\phi$, the transmission time $t$, and the weight of each photon $W$. For sources with narrow emission aperture, the initial attributes of each photon can be considered as $(x,y,z)=(0,0,0)$, $t=0$, and $W=1$. While propagating through the channel, each photon may interact with suspended particles; therefore, after each interaction, the photons' attributes should be updated according to the detailed steps described in \cite{tang2014impulse,cox2012simulation}. The tracking of each photon, i.e., updating its attributes, should be continued until the photon reaches the receiving plane (located at the distance $z=d_0$ from the transmitter, which $d_0$ is the link length) or its weight lies below a certain threshold $W_{th}$. Finally, those photons that reach the receiving plane with an acceptable weight will be selected as the detected photons if they are within the receiver aperture and their zenith angle is smaller than half of the receiver field of view (FOV).
The above steps must be repeated for all photons and all attributes of detected photons should be recorded to obtain the histogram of the received intensity versus time. The latter, which corresponds to the channel impulse response, can be obtained by summing the weight of detected photons within a specific propagation time (corresponding to the time interval of a specific bin of the histogram) and normalizing the sum by the total transmit weight.

In this paper, we study both collimated and diffusive UVLC links based on lasers and LEDs, respectively. It is worth mentioning that the main difference of channel FFIR  simulation for collimated and diffusive links is in the initial transmission direction, i.e., in the initial values of $\theta$ and $\phi$ for the emitted photons of the transmitter. Similar to \cite{cox2014simulating,cox2012simulation}, we assume Gaussian narrow beam lasers with the half divergence angle of $\theta_{div}$ and beam waist radius of $W_r$. As it is elaborated in \cite{cox2012simulation,leathers2004monte}, the emitted photons' initial zenith angle for collimated Gaussian beam lasers can be chosen as;
\begin{align}\label{theta_0,laser}
\theta^{\rm Collimated}_0=\theta_{div}\times\sqrt{-\ln\left(1-{{r}}_{\theta_0}^{\rm Collimated}\right)},
\end{align}
where ${r}_{\theta_0}^{\rm Collimated}\sim\boldsymbol{\rm U}(0,1)$ is a random variable (RV) with uniform distribution in the interval $[0,1]$. Also, we assume the beam is radially symmetric; hence, we randomly choose the initial value of the azimuth angle, $\phi_0^{\rm Collimated}$, with uniform distribution in the interval $[0,2\pi]$.

On the other hand, many practical underwater optical communication systems use LEDs as optical sources, due to their low cost, ease of engineering, and reduced difficulties in pointing and tracking \cite{cox2014simulating}. Similar to \cite{cox2012simulation,cox2014simulating}, we model diffusive LED-based links as generalized Lambertian radiant intensity light sources. For such sources the azimuthally symmetric polar angle distribution of radiant intensity can be approximated as;
\begin{align}\label{si_0}
\Psi_0(\theta)=\frac{m+1}{2\pi}\cos^m(\theta),
\end{align}
in which $m$ relates to the transmitter's semi-angle at half-power, $\theta_{1/2}$, as $\cos^m(\theta_{1/2})=0.5$, and $\Psi_0(\theta)$ is normalized such that $2\pi\int_{0}^{\pi}\Psi_0(\theta)\sin(\theta)d\theta=1$. Equating the integral of $2\pi\int_{0}^{\theta^{\rm Diffusive}_0}\Psi_0(\theta)\\\sin(\theta)d\theta$ to a uniformly chosen RV in the interval $[0,1]$, ${r}_{\theta_0}^{\rm Diffusive}$, yields the random value of the initial zenith angle as;
\begin{align}\label{theta_0^diffusive}
\theta^{\rm Diffusive}_0=\cos^{-1}\left(\sqrt[m+1]{1-{r}_{\theta_0}^{\rm Diffusive}}\right).
\end{align}
Once again, we assume the source is azimuthally symmetric; hence, we choose the initial value of the emitted photons' azimuth angle with a uniform distribution in the interval $[0,2\pi]$.
\subsection{Turbulence Effects of UVLC Channels}
While multipath reflection is the most significant cause of fading in acoustic and radio frequency (RF) links, the refractive index random variation of the medium conveying optical signals, also called as optical turbulence, is the main inducement of optical signals fading.
These random variations in underwater medium dominantly result from fluctuations in temperature and salinity \cite{tang2013temporal}.
To characterize turbulence effects, we multiply $h_{0,ij}\left(t\right)$ by a positive multiplicative fading coefficient, ${\tilde{h}}_{ij}$ \cite{navidpour2007ber,andrews2001laser,zhu2002free,karimi2009ber}.\footnote{Note that based on the numerical and experimental results presented in \cite{tang2013temporal} and \cite{jamali2016statistical}, respectively, the channel coherence time is on the order of $10^{-5}$ to $10^{-2}$ seconds, which is much larger than the channel typical delay spread values provided in \cite{tang2014impulse} (i.e., mainly smaller than tens of \si{ns}). Therefore, we can assume that conditioned on the fading coefficient value, $\tilde{h}_{i,j}$, the channel FFIR is $\tilde{h}_{i,j}h_{0,ij}(t)$, i.e., the total aggregated channel impulse response can in general be modeled as $\tilde{h}_{i,j}h_{0,ij}(t)$, with $\tilde{h}_{i,j}$ as a RV.} Weak oceanic turbulence can be modeled with lognormal distribution \cite{gerccekciouglu2014bit,yi2015underwater} as;
 \begin{align} \label{pdf lognormal}
\!\!f_{{\tilde{h}}_{ij}}({\tilde{h}}_{ij} )\!=\!\frac{1}{2{{\tilde{h}}_{ij}}\sqrt{2\pi {\sigma }^2_{X_{ij}}}}{\rm exp}\!\left(\!-\frac{{\left({{\rm ln}  ({{\tilde{h}}_{ij}})}\!-\!2{\mu }_{X_{ij}}\right)}^2}{8{\sigma }^2_{X_{ij}}}\right)\!,
 \end{align}
 where ${\mu }_{X_{ij}}$ and ${\sigma }_{X_{ij}}$ are respectively the mean and variance of the Gaussian distributed log-amplitude factor $X_{ij}=\frac{1}{2}{\rm ln}({{\tilde{h}}_{ij}})$.
 To ensure that fading neither amplifies nor attenuates the average power, we normalize fading amplitude such that $\E[{{{\tilde{h}}_{ij}} }]=1$, which implies that ${\mu }_{X_{ij}}=-{\sigma}^2_{X_{ij}}$ \cite{navidpour2007ber}.

 The scintillation index of a light wave with intensity $I_{ij}={{\tilde{h}}_{ij}}I_{0,ij}$ is defined by \cite{andrews2001laser};
 \begin{align} \label{S.I.}
 {\sigma^2_{I_{ij}}}=\frac{\E[{I^2_{ij}}]-\E^2[{I_{ij}}]}{\E^2[{I_{ij}}]}=\frac{\E[{{\tilde{h}}^2_{ij}}]-\E^2[{{\tilde{h}}_{ij}}]}{\E^2[{{\tilde{h}}_{ij}}]},
 \end{align}
in which $I_{0,ij}$ is the fading-free intensity. It can be shown that for the turbulent channel with lognormal fading distribution, the scintillation index is related to the log-amplitude variance as $\sigma^2_{I_{ij}}={\rm exp}(4{\sigma}^2_{X_{ij}})-1$ \cite{andrews2001laser}.
 Based on the numerical results presented in \cite{korotkova2012light,ata2014scintillations}, depending on various turbulence parameters, strong turbulence (specified by $\sigma^2_{I_{ij}}>1$ \cite{andrews2001laser}) can occur at distances as long as $100$ \si{m} and as short as $10$ \si{m}, which impressively differs from atmospheric channels where strong turbulence distances are on the order of kilometers \cite{korotkova2012light}. Therefore, mitigating such a strong turbulence demands advanced transmission methods like MIMO technique.
\subsection{System Model}
\begin{figure}
\centering
\includegraphics[width=3.4in]{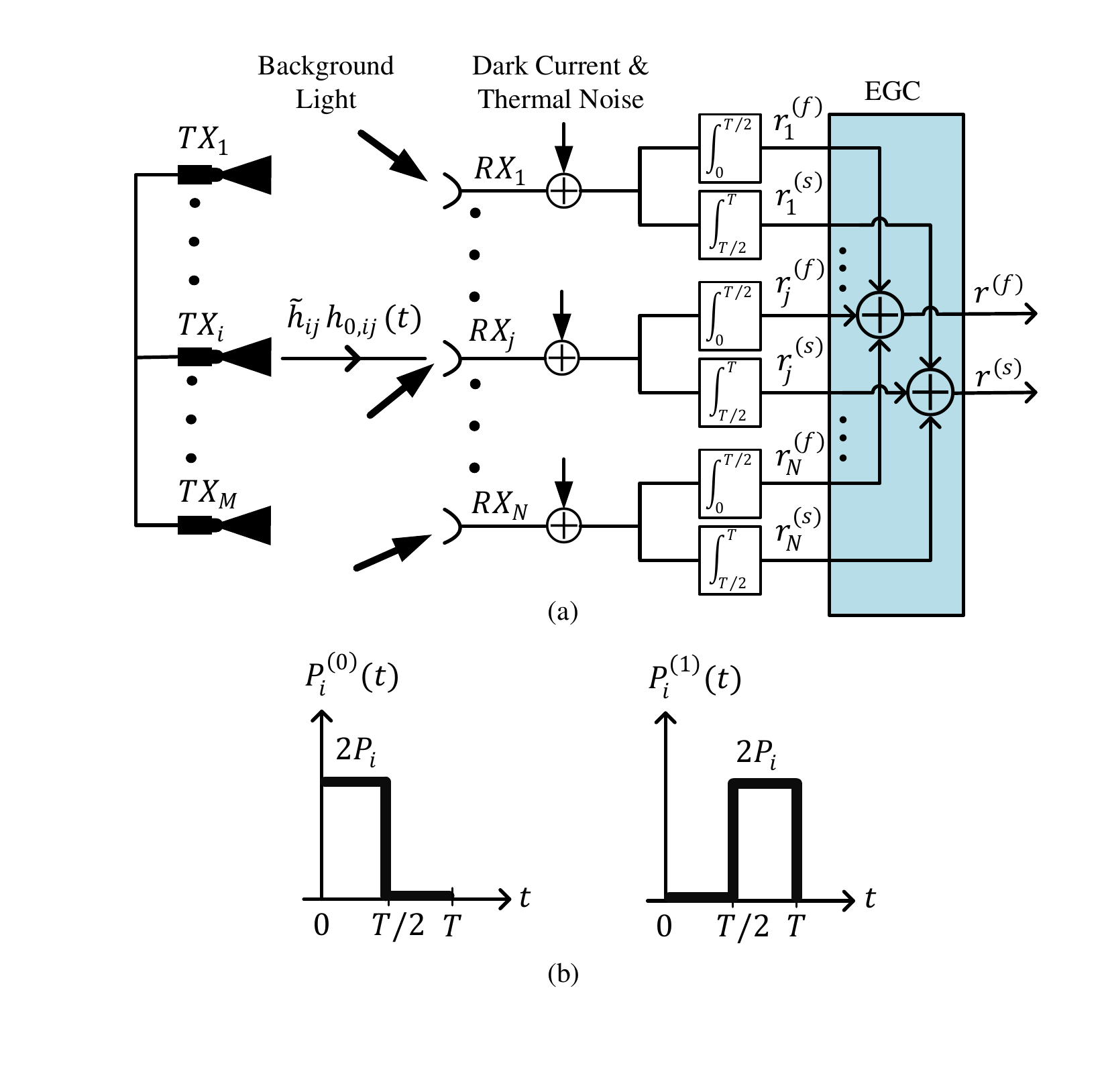}
\caption{(a) Block diagram of the proposed MIMO-UVLC system with BPPM signaling and EGC; (b) BPPM pulse shapes for the transmitted data bits ``$0$" and ``$1$".}
\vspace{-0.5cm}
\end{figure}
 We consider a UVLC system where the information signal is transmitted by $M$ transmitters, received by $N$ apertures, and combined using EGC. As it is depicted in Fig. 1(a), optical signal through propagation from the $i$th transmitter ${TX}_i$ to the $j$th receiver ${RX}_j$ experiences the aggregated channel impulse response $h_{i,j}(t)=\tilde{h}_{ij}h_{0,ij}(t)$, as discussed in the previous subsection.
 We assume intensity modulation direct-detection (IM/DD) with BPPM signaling. In this scheme, bits ``$0$" and ``$1$" of each time slot will be transmitted with pulse shapes $P^{(0)}_{i}(t)$ and $P^{(1)}_{i}(t)$, respectively, as it is shown in Fig. 1(b).
  The $i$th transmitter pulse carries the average transmitted power per bit of $P_{i}$
(or equivalently the peak transmitted power of $2P_{i}$), which relates to the total average transmitted power as $P=\sum_{i=1}^{M}P_{i}$.
Therefore, the $i$th transmitter data stream can be represented as;
\begin{align} \label{S_{i,j}}
S_{i}(t)=\sum^{\infty }_{k=-\infty }{{\overline{b_k}}P^{(0)}_{i}(t-kT)}+{\sum^{\infty }_{k=-\infty }{{b_k}P^{(1)}_{i}(t-kT)}},
\end{align}
where $b_k$ is the $k$th slot transmitted bit, $\overline{b_k}=1-b_k$ interprets the binary complement of $b_k$, and $T$ denotes the bit duration time. Eq. \eqref{S_{i,j}} implies that if $b_k=0$ the $k$th slot data will be transmitted with pulse shape $P^{(0)}_{i}(t-kT)$ and vice versa. Hence, we can express the received optical signal from $TX_i$ to $RX_j$ as;
\begin{align} \label{y_{i,j}}
& y_{i,j}\left(t\right)=S_{i}\left(t\right)*{\tilde{h}}_{ij}h_{0,ij}(t)=\nonumber\\
&~~~{\tilde{h}}_{ij}\!\!\sum_{k=-\infty}^{\infty}\!\!{{\overline{b_k}}{\Gamma}^{(0)}_{i,j}(t-kT)}+{\tilde{h}}_{ij}\!\!\sum_{k=-\infty}^{\infty}\!\!{{{b_k}}{\Gamma}^{(1)}_{i,j}(t-kT)},
\end{align}
in which $\Gamma^{(l)}_{i,j} \left(t\right)=h_{0,ij}(t)*P^{(l)}_{i}(t),~l=0,1$ and $*$ denotes the convolution operator. Furthermore, due to the photons' multiple scattering, portions of the transmitted signals of all transmitters can be captured by each $j$th receiver, i.e., $y_j\left(t\right)=\sum^M_{i=1}{y_{i,j}\left(t\right)}$ \cite{navidpour2007ber}.
\section{Analytical Study of the BER}
In this section, we analytically evaluate the system BER for both SISO and MIMO configurations with EGC. Our analytical approach in this section is based on additive white Gaussian noise (AWGN) model which, similar to \cite{navidpour2007ber} and \cite{lee2004part}, assumes the incoming optical signal as a constant coefficient (conditioned on the fading coefficient) and all of the noise components as an aggregated additive Gaussian noise component. We also assume that the signal-dependent shot noise has a negligible effect on the system performance compared to the other noise components \cite{jamali2015ber}; therefore, the combined noise variance is independent of the incoming optical signal power.
\subsection{SISO-UVLC}
Based on \eqref{y_{i,j}}, the integrated current over the first and second half of the $0$th bit duration time, when ``$b_0$" is sent at the $0$th time slot, can respectively be expressed as;
\begin{align}
r_{\rm SISO}^{(f,b_0)}=\tilde{h}\bigg[&\overline{b_0}\gamma^{(f,S_0)}+b_0\gamma^{(f,S_1)}+\nonumber\\
&\sum_{k=-L}^{-1}\left(\overline{b_k}\gamma^{(f,I_0,k)}+b_k\gamma^{(f,I_1,k)}\right)\bigg]+v^{(f)}_{T/2},\label{r_f}\\
r_{\rm SISO}^{(s,b_0)}=\tilde{h}\bigg[&\overline{b_0}\gamma^{(s,S_0)}+b_0\gamma^{(s,S_1)}+\nonumber\\
&\sum_{k=-L}^{-1}\left(\overline{b_k}\gamma^{(s,I_0,k)}+b_k\gamma^{(s,I_1,k)}\right)\bigg]+v^{(s)}_{T/2},\label{r_s}
\end{align}
in which $\gamma^{(f,S_l)}={\mathcal{R}}\int_{0}^{T/2}\Gamma^{(l)}(t)dt$, and $\gamma^{(s,S_l)}={\mathcal{R}}\int_{T/2}^{T}\Gamma^{(l)}(t)dt$, $l=0,1$. Also $L$ is the channel memory and ${\mathcal{R}}\!=\!\eta q/hf$ is the photodetector resposivity, where $\eta$ is the photodetector quantum efficiency, $q=1.602\times 10^{-19}$ C is the electron charge, $h=6.626\times10^{-34}~{\rm J.s}$ is Planck's constant, and $f$ is the optical source frequency. Moreover, $\gamma^{(f,I_l,k)}={\mathcal{R}}\int_{0}^{T/2}\Gamma^{(l)}(t-kT)dt$, and $\gamma^{(s,I_l,k)}={\mathcal{R}}\int_{T/2}^{T}\Gamma^{(l)}(t-kT)dt$, $l=0,1$. Obviously, $\gamma^{(f,S_0)}=\gamma^{(s,S_1)}$ and $\gamma^{(f,I_0,k)}=\gamma^{(s,I_1,k)}$. In addition, $v^{(f)}_{{T}/{2}}$ and $v^{(s)}_{{T}/{2}}$ are the integrated noise components of the first and second half of the bit duration time, each with a Gaussian distribution with mean zero and variance $\sigma^2_{T_c}={4K_bT_eBT_c^2}/{R_L}+2q{\mathcal{R}}P_{BG}BT_c^2+2qI_{dc}BT_c^2$, where $T_c=T/2$ is the chip duration time. $K_b$, $T_e$, $B$, and $R_L$ are Boltzmann's constant, the receiver equivalent temperature, electronic bandwidth, and load resistance, respectively. Moreover, $P_{BG}$ is the received background power and $I_{dc}$ is the photodetector dark current \cite{jazayerifar2006atmospheric,lee2004part,giles2005underwater}.
 It is worth mentioning that the large coherence time of the channel compared to the bit duration time implies to the same fading coefficient for thousands up to millions of consecutive bits \cite{tang2013temporal}. Therefore, we adopted the same fading coefficient for all consecutive bits in \eqref{r_f} and \eqref{r_s}.

In this section, we assume bit-by-bit detection which is suboptimal in the presence of ISI \cite{einarsson2008principles}. In this case, the receiver simply compares its integrated current over the first and second half of each time slot to detect the received data bits. Then the conditional probability of error on the transmitted bit ``$0$" can be determined as;
\begin{align}\label{pe|0}
\!P^{\rm SISO}_{be|0,\tilde{h},b_k}\!\!&=\Pr\left(r^{(f,b_0)}_{\rm SISO}\leq r^{(s,b_0)}_{\rm SISO}|b_0=0,\tilde{h},b_k\right)\nonumber\\
&=\!\Pr\!\left(\!v^{(s)}_{T/2}\!-\!v^{(f)}_{T/2}\!\geq\! \tilde{h}\!\left[\gamma^{(f,S_0)}\!-\!\gamma^{(s,S_0)}\!+\!\!\!\!\sum_{k=-L}^{-1}\!\!\!\!C^{(k)}\right]\!\right)\nonumber\\
& =Q\!\!\left(\!\!\frac{\tilde{h}}{\sqrt{2\sigma^2_{T_c}}}\!\left[\gamma^{(f,S_0)}-\gamma^{(s,S_0)}+\!\!\!\sum_{k=-L}^{-1}\!\!\!C^{(k)}\right]\!\right)\!,
\end{align}
where $Q\left(x\right)=({1}/{\sqrt{2\pi }})\int^{\infty }_x{{\rm exp}({-{y^2}/{2}})}dy$ is the Gaussian-Q function and $C^{(k)}={b_k}\left[\gamma^{(f,I_1,k)}-\gamma^{(s,I_1,k)}\right]+\overline{b_k}\left[\gamma^{(f,I_0,k)}-\gamma^{(s,I_0,k)}\right]$. Similar steps result to the following expression for the conditional probability of error on the transmitted bit ``$1$";
\begin{align}\label{pe|1}
\!P^{\rm SISO}_{be|1,\tilde{h},b_k}\!\!=Q\!\!\left(\!\!\frac{\tilde{h}}{\sqrt{2\sigma^2_{T_c}}}\!\left[\gamma^{(s,S_1)}-\gamma^{(f,S_1)}-\!\!\!\sum_{k=-L}^{-1}\!\!\!C^{(k)}\right]\!\right)\!.
\end{align}
The final BER can then be obtained by averaging over the fading coefficient and all $2^L$ possible data sequences for the transmitted data bits as;
\begin{align}\label{SecIII-A-5}
P^{\rm SISO}_{be}=\frac{1}{2^L}\sum_{b_k}\int_{0}^{\infty}\frac{1}{2}\left[P^{\rm SISO}_{be|0,\tilde{h},b_k}+P^{\rm SISO}_{be|1,{\tilde{h}},b_k}\right]f_{{\tilde{h}}}(\tilde{h})d{\tilde{h}}.
\end{align}

In the special case of weak oceanic turbulence, where fading coefficient can be modeled as a lognormal RV \cite{gerccekciouglu2014bit,yi2015underwater}, the averaging over fading coefficient can effectively be calculated using Gauss-Hermite quadrature formula [30, Eq. (25.4.46)] as follows;
\begin{align} \label{SecIII-A-6}
P^{\rm SISO}_{be|b_0,b_k}&=\int_{0}^{\infty}P^{\rm SISO}_{be|b_0,{\tilde{h}},b_k}f_{{\tilde{h}}}(\tilde{h})d{\tilde{h}}\nonumber\\
&\approx\frac{1}{\sqrt{\pi}}\sum_{q=1}^{V}w_qP^{\rm SISO}_{be|b_0,{\tilde{h}}=\exp\left(2x_q\sqrt{2\sigma^2_X}+2\mu_X\right),b_k},
\end{align}
in which $V$ is the order of approximation, $w_q,~q=1,2,...,V$, are weights of the $V$th-order approximation, and $x_q$ is the $q$th zero of the $V$th-order Hermite polynomial, $H_V(x)$ \cite{navidpour2007ber,abramowitz1970handbook}.
\subsection{MIMO-UVLC}
Based on Eq. \eqref{y_{i,j}} and the receiver structure in Fig. 1(a), corresponding to the MIMO UVLC system with EGC\footnote{It is shown in \cite{jamali2015performanceMIMO} that EGC provides a close performance to the optimal combiner, and hence, due to its lower complexity, can be consider as a practical combining method in MIMO UVLC systems.}, we can express the receiver integrated current over the first and second half of the $0$th bit duration time, when ``$b_0$" is sent at the $0$th time slot, respectively as;
\begin{align}
& r_{\rm MIMO}^{(f,b_0)}=\sum_{i=1}^{M}\sum_{j=1}^{N}{\tilde{h}}_{ij}\left(\overline{b_0}\gamma_{i,j}^{(f,S_0)}+b_0\gamma_{i,j}^{(f,S_1)}\right)\nonumber\\
&+\!\!\sum_{i=1}^{M}\sum_{j=1}^{N}{\tilde{h}}_{ij}\!\!\!\!\!\sum_{k=-L_{ij}}^{-1}\!\!\!\!\left(\overline{b_k}\gamma_{i,j}^{(f,I_0,k)}\!\!+\!b_k\gamma_{i,j}^{(f,I_1,k)}\right)\!+\!\sum_{j=1}^{N}\!v^{(f,j)}_{T/2},\label{r^MIMO_f}\\
& r_{\rm MIMO}^{(s,b_0)}=\sum_{i=1}^{M}\sum_{j=1}^{N}{\tilde{h}}_{ij}\left(\overline{b_0}\gamma_{i,j}^{(s,S_0)}+b_0\gamma_{i,j}^{(s,S_1)}\right)\nonumber\\
&+\!\!\sum_{i=1}^{M}\sum_{j=1}^{N}{\tilde{h}}_{ij}\!\!\!\!\!\sum_{k=-L_{ij}}^{-1}\!\!\!\!\left(\overline{b_k}\gamma_{i,j}^{(s,I_0,k)}\!\!+\!b_k\gamma_{i,j}^{(s,I_1,k)}\right)\!+\!\sum_{j=1}^{N}\!v^{(s,j)}_{T/2},\label{r^MIMO_s}
\end{align}
where $L_{ij}$ is the memory of the channel between the $i$th transmitter and $j$th receiver. $v^{(f,j)}_{{T}/{2}}$ and $v^{(s,j)}_{{T}/{2}}$ are the integrated noise components of the first and second half of the bit duration time, each with a Gaussian distribution with mean zero and variance $\sigma^2_{T_c}$ \cite{lee2004part}. Moreover, $\gamma_{i,j}^{(f,S_l)}={\mathcal{R}}\int_{0}^{T/2}\Gamma_{i,j}^{(l)}(t)dt$, $\gamma_{i,j}^{(s,S_l)}={\mathcal{R}}\int_{T/2}^{T}\Gamma_{i,j}^{(l)}(t)dt$, $\gamma_{i,j}^{(f,I_l,k)}={\mathcal{R}}\int_{0}^{T/2}\Gamma_{i,j}^{(l)}(t-kT)dt$, and $\gamma_{i,j}^{(s,I_l,k)}={\mathcal{R}}\int_{T/2}^{T}\Gamma_{i,j}^{(0)}(t-kT)dt$, $l=0,1$.

The symbol-by-symbol receiver simply compares $r_{\rm MIMO}^{(f,b_0)}$ and $r_{\rm MIMO}^{(s,b_0)}$ to detect the received data bits. Therefore, the conditional probabilities of error on the transmitted bits ``$0$" and ``$1$" can respectively be calculated as;
\begin{align}
& P^{\rm MIMO}_{be|0,\tilde{\boldsymbol{H}},b_k}\!\!\!=\!Q\!\left(\!\sum_{i=1}^{M}\!\sum_{j=1}^{N}\!\frac{{\tilde{h}}_{ij}}{\sqrt{2N\sigma^2_{T_c}}}\!\!\left[\!\gamma_{ij}^{(f,S_0)}\!\!-\!\gamma_{ij}^{(s,S_0)}\!\!+\!\!\!\!\!\!\sum_{k=-L_{ij}}^{-1}\!\!\!\!\!C_{i,j}^{(k)}\!\right]\!\right)\!\!,\label{SecIII-B-3}\\
& P^{\rm MIMO}_{be|1,\tilde{\boldsymbol{H}},b_k}\!\!\!=\!Q\!\left(\!\sum_{i=1}^{M}\!\sum_{j=1}^{N}\!\frac{{\tilde{h}}_{ij}}{\sqrt{2N\sigma^2_{T_c}}}\!\!\left[\!\gamma_{ij}^{(s,S_1)}\!\!-\!\gamma_{ij}^{(f,S_1)}\!\!-\!\!\!\!\!\!\sum_{k=-L_{ij}}^{-1}\!\!\!\!\!C_{i,j}^{(k)}\!\right]\!\right)\!\!,\label{SecIII-B-4}
\end{align}
in which $\tilde{\boldsymbol{H}}=({\tilde{h}}_{11},{\tilde{h}}_{12},...,{\tilde{h}}_{MN})$ is the fading coefficients vector, and $C_{i,j}^{(k)}={b_k}\left[\gamma_{i,j}^{(f,I_1,k)}-\gamma_{i,j}^{(s,I_1,k)}\right]+\overline{b_k}\left[\gamma_{i,j}^{(f,I_0,k)}-\gamma_{i,j}^{(s,I_0,k)}\right]$. Then if the maximum channel memory is $L_{\rm max}={\rm max}\{L_{11},L_{12},...,L_{MN}\}$, the final BER can be obtained by averaging over the fading coefficients vector and all $2^{L_{\rm max}}$ possible data sequences for the transmitted bits as\footnote{Note that if some of the links have smaller channel memory than $L_{\rm max}$, i.e., if $L_{ij}<L_{\rm max}$, we can consider additional zero coefficients as $\gamma_{i,j}^{(.,.,k)}=0,~-L_{\rm max}\leq k\leq -L_{ij}-1$ for those channels to make all the channels with the same memory $L_{\rm max}$.};
\begin{align}\label{SecIII-B-5}
P^{\rm MIMO}_{be}\!=\!\frac{1}{2^{L_{\rm max}}}\!\sum_{b_k}\!\int_{\tilde{\boldsymbol{H}}}\!\frac{1}{2}\!\left[P^{\rm MIMO}_{be|0,\tilde{\boldsymbol{H}},b_k}\!\!+\!P^{\rm MIMO}_{be|1,{\tilde{\boldsymbol{H}}},b_k}\right]\!f_{{\tilde{\boldsymbol{H}}}}(\tilde{\boldsymbol{H}})d{\tilde{\boldsymbol{H}}},
\end{align}
where $f_{{\tilde{\boldsymbol{H}}}}(\tilde{\boldsymbol{H}})$ is the joint probability density function (PDF) of fading coefficients in $\tilde{\boldsymbol{H}}$.

 It is worth mentioning that in the special case of lognormal fading, the ($M\times N$)-dimensional averaging integral of $\int_{\tilde{\boldsymbol{H}}}P^{\rm MIMO}_{be|b_0,\tilde{\boldsymbol{H}},b_k}f_{{\tilde{\boldsymbol{H}}}}(\tilde{\boldsymbol{H}})d{\tilde{\boldsymbol{H}}}$ can effectively be calculated by ($M\times N$)-dimensional finite series, using Gauss-Hermite quadrature formula, as follows \cite{jamali2015performanceMIMO};
\begin{align}\label{SecIII-B-6}
&P^{\rm MIMO}_{be|b_0,b_k}=\int_{\tilde{\boldsymbol{H}}}P^{\rm MIMO}_{be|b_0,\tilde{\boldsymbol{H}},b_k}f_{{\tilde{\boldsymbol{H}}}}(\tilde{\boldsymbol{H}})d{\tilde{\boldsymbol{H}}}\approx\frac{1}{{\pi}^{M\times N/2}}\sum_{q_{11}=1}^{V_{11}}w^{(11)}_{q_{11}}\nonumber\\&\times\!\!\!\sum_{q_{21}=1}^{V_{21}}\!\!\!w^{(21)}_{q_{21}}...\!\!\!\!\!\sum_{q_{MN}=1}^{V_{MN}}\!\!\!\!w^{(MN)}_{q_{MN}}\!\times\! P^{\rm MIMO}_{be|b_0,{\tilde{h}}_{ij}=\exp\left(\!2x^{(ij)}_{q_{ij}}\!\!\sqrt{2\sigma^2_{X_{ij}}}+2\mu_{X_{ij}}\!\right),b_k}.
\end{align}
Moreover, based on \eqref{SecIII-B-3} and \eqref{SecIII-B-4}, we can reformulate $P^{\rm MIMO}_{be|b_0,\tilde{\boldsymbol{H}},b_k}$ as $P^{\rm MIMO}_{be|b_0,\tilde{\boldsymbol{H}},b_k}=Q\left(\sum_{i=1}^{M}\sum_{j=1}^{N}G_{i,j}^{(b_0)}{\tilde{h}}_{ij}\right)$ where;
\begin{align}
G_{i,j}^{(0)}=\frac{1}{\sqrt{2N\sigma^2_{T_c}}}\left[\gamma_{ij}^{(f,S_0)}-\gamma_{ij}^{(s,S_0)}+\sum_{k=-L_{ij}}^{-1}C_{i,j}^{(k)}\right],\label{SecIII-B-7}\\
G_{i,j}^{(1)}=\frac{1}{\sqrt{2N\sigma^2_{T_c}}}\left[\gamma_{ij}^{(s,S_1)}-\gamma_{ij}^{(f,S_1)}-\sum_{k=-L_{ij}}^{-1}C_{i,j}^{(k)}\right].\label{SecIII-B-8}
\end{align}
In the special case of lognormal fading, we can approximate the weighted sum of $\sum_{i=1}^{M}\sum_{j=1}^{N}G_{i,j}^{(b_0)}{\tilde{h}}_{ij}$ as an equivalent lognormal RV, $\alpha^{(b_0)}=\exp(2z^{(b_0)})$, with the following log-amplitude mean and variance \cite{safari2008relay};
\begin{align} \label{mu}
      \mu_{z^{(b_0)}}=\frac{1}{2}{\rm ln}\bigg(\sum_{j=1}^N\sum_{i=1}^MG^{(b_0)}_{i,j}\bigg)-\sigma^2_{z^{(b_0)}},
\end{align}
\begin{align} \label{sigma2}
\sigma^2_{z^{(b_0)}}=\frac{1}{4}{\rm ln}\left(1+\frac{\sum_{j=1}^N\sum_{i=1}^M\left(G^{(b_0)}_{i,j}\right)^2\left(e^{4\sigma^2_{X_{ij}}}-1\right)}{\left(\sum_{j=1}^N\sum_{i=1}^MG^{(b_0)}_{i,j}\right)^2}\right).
\end{align}
Therefore, using the aforementioned approximation, the ($M\times N$)-dimensional integrals in \eqref{SecIII-B-5} reduce to one-dimensional counterparts of the form $\int_{\alpha^{(b_0)}}P^{\rm MIMO}_{be|b_0,\alpha^{(b_0)},b_k}f_{\alpha^{(b_0)}}(\alpha^{(b_0)})d\alpha^{(b_0)}$, where $f_{\alpha^{(b_0)}}(\alpha^{(b_0)})$ is the PDF of the equivalent lognormal RV.
 \section{BER Evaluation Using Photon-Counting Method}
In this section, we apply saddle-point approximation \cite{einarsson2008principles}, which is based on photon-counting method, to evaluate the system BER in the presence of shot noise. Specifically, in order to take into account the shot noise effect, in this section, we consider the incoming optical signal as a Poisson distributed RV (conditioned on $\tilde{\boldsymbol{H}}$). Moreover, we consider thermal noise with Gaussian distribution, and dark current and background light both with Poisson distribution \cite{einarsson2008principles}.
The bit-by-bit photon-counting receiver compares the photoelectrons count over the first and second half of the bit duration time to detect the received data. Then the conditional probability of error, when ``$b_0$" is sent, can be obtained using saddle-point approximation as \cite{einarsson2008principles,karimi2009ber};
\begin{align} \label{saddlepoint}
P_{be|b_0,\tilde{\boldsymbol{H}},b_k}\!\!\!=\!\Pr\!\left(\!A^{(b_0)}\!\geq\! 0{\big|}b_0,\tilde{\boldsymbol{H}},b_k\right)\!\!=\!\frac{{\rm exp\!}\left[{\Phi }_{A^{(b_0)}|\tilde{\boldsymbol{H}},b_k}\!(s_{b_0})\right]}{\sqrt{2\pi {\Phi }^{''}_{A^{(b_0)}|\tilde{\boldsymbol{H}},b_k}\!(s_{b_0})}},
\end{align}
in which, $A^{(0)}=u^{(s,0)}-u^{(f,0)}$ and $A^{(1)}=u^{(f,1)}-u^{(s,1)}$, where $u^{(f,b_0)}$ and $u^{(s,b_0)}$ are respectively the photoelectrons count of the first and second half of the bit duration time when ``$b_0$" is sent.
    Moreover, $s_{b_0}$ is the positive and real root of the first derivative of ${\Phi }_{A^{(b_0)}|\tilde{\boldsymbol{H}},b_k}(s)$, i.e., ${\Phi }^{'}_{A^{(b_0)}|\tilde{\boldsymbol{H}},b_k}(s_{b_0})=0$ while ${\Phi }_{A^{(b_0)}|\tilde{\boldsymbol{H}},b_k}(s)$ itself is defined as;
    \begin{align} \label{Phi_{A|0,H}}
    {\Phi }_{A^{(b_0)}|\tilde{\boldsymbol{H}},b_k}(s)={\rm ln}\left[\Psi_{A^{(b_0)}|\tilde{\boldsymbol{H}},b_k}(s)\right]-{\rm ln}|s|,
    \end{align}
    where $\Psi_{A^{(b_0)}|\tilde{\boldsymbol{H}},b_k}(s)$ is the moment generating function (MGF) of $A^{(b_0)}$ conditioned on $\tilde{\boldsymbol{H}}$ and $\{b_k\}_{k=-L_{\rm max}}^{-1}$, defined as $\E[e^{sA^{(b_0)}}|\tilde{\boldsymbol{H}},b_k]$, which for $b_0=0$ and $1$ can respectively be calculated as \cite{karimi2009ber};
\begin{align}
&\Psi_{A^{(0)}|\tilde{\boldsymbol{H}},b_k}(s)=\Psi_{u^{(s,0)}|\tilde{\boldsymbol{H}},b_k}(s)\times\Psi_{u^{(f,0)}|\tilde{\boldsymbol{H}},b_k}(-s),\label{P_{A=|0,H}}\\
&\Psi_{A^{(1)}|\tilde{\boldsymbol{H}},b_k}(s)=\Psi_{u^{(f,1)}|\tilde{\boldsymbol{H}},b_k}(s)\times\Psi_{u^{(s,1)}|\tilde{\boldsymbol{H}},b_k}(-s),\label{P_{A=|1,H}}
    \end{align}
in which $\Psi_{u^{(.,.)}|\tilde{\boldsymbol{H}},b_k}(s)$ is the MGF of $u^{(.,.)}$ conditioned on $\tilde{\boldsymbol{H}}$ and $\{b_k\}_{k=-L_{\rm max}}^{-1}$. 
In this section, we extract MGFs of the receiver output for both SISO and MIMO configurations to obtain the conditional BERs through \eqref{saddlepoint}-\eqref{P_{A=|1,H}}. The final BER $P_{be}$ can then be measured by averaging over $\tilde{\boldsymbol{H}}$ and $\{b_k\}_{k=-L_{\rm max}}^{-1}$.
\subsection{SISO UVLC Link}
In SISO scheme, the photo-detected signal generated by the integrate-and-dump circuit can be expressed as $u_{\rm SISO}^{(f,b_0)}=y_{\rm SISO}^{(f,b_0)}+v^{(f)}_{th}$ and $u_{\rm SISO}^{(s,b_0)}=y_{\rm SISO}^{(s,b_0)}+v^{(s)}_{th}$ for the first and second half of the bit duration time, respectively. $v^{(f)}_{th}$ and $v^{(s)}_{th}$ are the integrated thermal noise components of the receiver over the first and second half of the bit duration time, respectively, each with a Gaussian distribution with mean zero and variance $\sigma^2_{th}={4K_bT_eBT^2_c}/{R_Lq^2}$. Moreover, $y_{\rm SISO}^{(f,b_0)}$ and $y_{\rm SISO}^{(s,b_0)}$ are respectively the photoelectrons count over the first and second half of the bit duration time, resulted from the incoming optical signal, dark current, and background light. Conditioned on $\tilde{h}$ and $\{b_k\}_{k=-L}^{-1}$, $y_{\rm SISO}^{(f,b_0)}$ and $y_{\rm SISO}^{(s,b_0)}$ are Poisson RVs with mean $m_{\rm SISO}^{(f,b_0)}$ and $m_{\rm SISO}^{(s,b_0)}$, respectively as;
 \begin{align}
& m_{\rm SISO}^{(f,b_0)}=\frac{\tilde{h}}{q}\bigg[\overline{b_0}\gamma^{(f,S_0)}+b_0\gamma^{(f,S_1)}\nonumber\\
&~~~+\sum_{k=-L}^{-1}\left(\overline{b_k}\gamma^{(f,I_0,k)}+b_k\gamma^{(f,I_1,k)}\right)\bigg]+m^{(bd)}_{\rm SISO},\label{m_f}\\
 & m_{\rm SISO}^{(s,b_0)}=\frac{\tilde{h}}{q}\bigg[\overline{b_0}\gamma^{(s,S_0)}+b_0\gamma^{(s,S_1)}\nonumber\\
 &~~~+\sum_{k=-L}^{-1}\left(\overline{b_k}\gamma^{(s,I_0,k)}+b_k\gamma^{(s,I_1,k)}\right)\bigg]+m^{(bd)}_{\rm SISO},\label{m_s}
 \end{align}
where $m^{(bd)}_{\rm{SISO}}$ is the mean of photoelectrons count, due to background light and dark current noise, integrated over $T_c$. Since Poisson distribution is assumed for these two noise components, the mean and variance of the aforementioned count process is the same, i.e., $m^{(bd)}_{\rm{SISO}}=\sigma^2_{bd,\rm{SISO}}=(n_b+n_d)T/2$, where $n_b=2\eta P_{BG}BT_c/hf$ and $n_d=2I_{dc}BT_c/q$ are the photoelectrons count rates due to background light and dark current noise, respectively.

Since $y_{\rm SISO}^{(f,b_0)}$ and $v^{(f)}_{th}$ are two independent RVs, MGF of their sum, $u_{\rm SISO}^{(f,b_0)}$, is the product of each one's MGF, i.e., 
\begin{align} \label{27}
\Psi_{u_{\rm SISO}^{(f,b_0)}|\tilde{h},b_k}(s)&=\Psi_{y_{\rm SISO}^{(f,b_0)}|\tilde{h},b_k}(s)\times\Psi_{v^{(f)}_{th}}(s)\nonumber\\
&={\rm exp}\left(\frac{\sigma^2_{th}}{2}s^2+m_{\rm SISO}^{(f,b_0)}(e^s-1)\right).
\end{align}
Also the MGF of the receiver integrated output over the second half of the bit duration time can be obtained as;
\begin{align} \label{28}
\Psi_{u_{\rm SISO}^{(s,b_0)}|\tilde{h},b_k}(s)={\rm exp}\left(\frac{\sigma^2_{th}}{2}s^2+m_{\rm SISO}^{(s,b_0)}(e^s-1)\right).
\end{align}
By substituting \eqref{27} and \eqref{28} in \eqref{saddlepoint}-\eqref{P_{A=|1,H}}, one can obtain the conditional probability of error $P^{\rm SISO}_{be|b_0,\tilde{h},b_k}$ using saddle-point approximation, and the final BER can then be evaluated similar to \eqref{SecIII-A-5}.
\subsection{MIMO UVLC Link}
Similar to previous subsection, the photo-detected signal generated by the integrate-and-dump circuit of the MIMO receiver can be expressed as $u_{\rm MIMO}^{(f,b_0)}=y_{\rm MIMO}^{(f,b_0)}+v^{(f,N)}_{th}$ and $u_{\rm MIMO}^{(s,b_0)}=y_{\rm MIMO}^{(s,b_0)}+v^{(s,N)}_{th}$ for the first and second half of the bit duration time, respectively, where $v^{(f,N)}_{th}$ and $v^{(s,N)}_{th}$ are zero mean Gaussian RVs with variance $N\sigma^2_{th}$, corresponding to the integrated and combined thermal noise components of the receiver over the first and second half of the bit duration time, respectively. Moreover, $y_{\rm MIMO}^{(f,b_0)}$ and $y_{\rm MIMO}^{(s,b_0)}$ are the photoelectrons count of the MIMO system over the first and second half of the bit duration time, respectively. Conditioned on $\tilde{\boldsymbol{H}}$ and $\{b_k\}_{k=-L_{\rm max}}^{-1}$, $y_{\rm MIMO}^{(f,b_0)}$ and $y_{\rm MIMO}^{(s,b_0)}$ are Poisson RVs with mean;
 \begin{align}
 & u_{\rm MIMO}^{(f,b_0)}=\sum_{i=1}^{M}\sum_{j=1}^{N}\frac{{\tilde{h}}_{ij}}{q}\left(\overline{b_0}\gamma_{i,j}^{(f,S_0)}+b_0\gamma_{i,j}^{(f,S_1)}\right)\nonumber\\
 &~+\!\!\sum_{i=1}^{M}\sum_{j=1}^{N}\!\frac{{\tilde{h}}_{ij}}{q}\!\!\!\sum_{k=-L_{ij}}^{-1}\!\!\!\!\left(\overline{b_k}\gamma_{i,j}^{(f,I_0,k)}\!\!+\!b_k\gamma_{i,j}^{(f,I_1,k)}\right)\!+\!m^{(bd)}_{\rm MIMO},\label{u^MIMO_f}\\
 & u_{\rm MIMO}^{(s,b_0)}=\sum_{i=1}^{M}\sum_{j=1}^{N}\frac{{\tilde{h}}_{ij}}{q}\left(\overline{b_0}\gamma_{i,j}^{(s,S_0)}+b_0\gamma_{i,j}^{(s,S_1)}\right)\nonumber\\
 &~+\!\!\sum_{i=1}^{M}\sum_{j=1}^{N}\!\frac{{\tilde{h}}_{ij}}{q}\!\!\!\sum_{k=-L_{ij}}^{-1}\!\!\!\!\left(\overline{b_k}\gamma_{i,j}^{(s,I_0,k)}\!\!+\!b_k\gamma_{i,j}^{(s,I_1,k)}\right)\!+\!m^{(bd)}_{\rm MIMO},\label{u^MIMO_s}
  \end{align}
respectively, in which $m^{(bd)}_{\rm MIMO}=(n_b+Nn_d)T/2$.\footnote{Note that the received background power is proportional to the receiver aperture area. Since, for the sake of fairness, we have assumed that the sum of the receiving apertures in MIMO scheme has the same area as the receiver aperture of SISO scheme, we can conclude that $n_b^{\rm MIMO}=n_b^{\rm SISO}=n_b$. Furthermore, each receiver in MIMO scheme produces a zero mean Gaussian distributed thermal noise with variance $\sigma^2_{th}$ and also a Poisson distributed dark current noise with count rate $n_d$.}
Eventually, the MGF of the receiver integrated output over the first and second half of the bit duration time can respectively be obtained as;
\begin{align}
&\!\Psi_{u_{\rm MIMO}^{(f,b_0)}|\tilde{\boldsymbol{H}},b_k}\!(s)\!={\rm exp}\left(\frac{N\sigma^2_{th}}{2}s^2+m_{\rm MIMO}^{(f,b_0)}(e^s-1)\right),\label{31}\\
&\!\Psi_{u_{\rm MIMO}^{(s,b_0)}|\tilde{\boldsymbol{H}},b_k}\!(s)\!={\rm exp}\left(\frac{N\sigma^2_{th}}{2}s^2+m_{\rm MIMO}^{(s,b_0)}(e^s-1)\right).\label{32}
\end{align}
By substituting \eqref{31} and \eqref{32} in \eqref{saddlepoint}-\eqref{P_{A=|1,H}}, the conditional probability of error $P^{\rm MIMO}_{be|b_0,\boldsymbol{H},b_k}$ can be obtained and the final BER can then be evaluated similar to \eqref{SecIII-B-5}. We should emphasize that the approximation to the sum of lognormal RVs can also be applied (similar to \eqref{SecIII-B-7}-\eqref{sigma2}) to approximate the ($M\times N$)-dimensional integrals with one-dimensional counterparts.
\section{Multiple-Symbol Detection}
The highly scattering nature of UVLC channels causes significant interference between the received symbols, especially for diffusive link geometries. Such a strong ISI for typical data rates deteriorates the performance of UVLC systems and demands intelligent reception algorithms. Therefore, in this section, we derive the (G)MSD algorithm(s) for SISO UVLC systems with BPPM signaling to simultaneously detect a block of consecutive bits and improve the system BER.

Based on Eq. \eqref{y_{i,j}}, the received time-domain signal can be expressed as;
\begin{align} \label{eq36}
y_{\rm rec}\left(t\right)=&{\mathcal{R}} {\tilde{h}}\sum_{k=-\infty}^{\infty}{{\overline{b_k}}{\Gamma}^{(0)}(t-kT)}+\nonumber\\
&{\mathcal{R}}{\tilde{h}}\sum_{k=-\infty}^{\infty}{{{b_k}}{\Gamma}^{(1)}(t-kT)}+Z(t),
\end{align}
where $Z(t)$ is the receiver noise component. Using the Karhunen-Lo\`{e}ve expansion, we can expand the received signal into a series of the form;
\begin{align} \label{eq37}
y_{\rm rec}\left(t\right)=\sum_{m=1}^{\infty}{{y_{{\rm rec},m}}{\varphi}_{m}(t)},
\end{align}
in which $\{{\varphi}_{m}(t)\}$ is a complete set of orthogonal functions, and ${y_{{\rm rec},m}}$ is the observable RV obtained by projecting $y_{\rm rec}(t)$ onto the set $\{{\varphi}_{m}(t)\}$, which can be obtained as;
\begin{align} \label{eq38}
& {y_{{\rm rec},m}}={\mathcal{R}}{\tilde{h}}\sum_{k=-\infty}^{\infty}{{\overline{b_k}}{\Gamma}^{(0)}_{m,k}}+
{\mathcal{R}}{\tilde{h}}\sum_{k=-\infty}^{\infty}{{{b_k}}{\Gamma}^{(1)}_{m,k}}+{Z_{m}},
\end{align}
where ${\Gamma}^{(0)}_{m,k} = \langle  {\Gamma}^{(0)}(t-kT).{\varphi}_{m}(t) \rangle$, $ {\Gamma}^{(1)}_{m,k} = \langle {\Gamma}^{(1)}(t-kT).{\varphi}_{m}(t)\rangle$, and $\langle .\rangle$ denotes the projection operator. Furthermore, ${Z_{m}}=\langle Z(t).{\varphi}_{m}(t)\rangle$ which has a Gaussian distribution with mean zero and variance $\sigma^2_{Z_{m}}$. The joint PDF of $F$ observable RVs, $\underline{y}_{\rm rec}^{(F)}\triangleq[{y_{{\rm rec},1}},{y_{{\rm rec},2}},...,{y_{{\rm rec},F}}]$, conditioned on the transmitted data sequence $\underline{b}_P\triangleq[b_{0},b_{1},...,b_{P-1}]$, where $P$ is the length of detection window, can be expressed as;
\begin{align} \label{eq39}
& f(\underline{y}_{\rm rec}^{(F)}{|}\underline{b}_P,\tilde{h})=\frac{1}{(2\pi{\sigma^2_{Z_{m}}})^{F/2}}\times \nonumber
\\&\exp\left(\frac{-1}{2{\sigma^2_{Z_{m}}}}{\sum_{m=1}^{F}{{\Big|}{y_{{\rm rec},m}}-{\mathcal{R}}{\tilde{h}}\sum_{k=0}^{P-1}{{\overline{b_k}}{\Gamma}^{(0)}_{m,k}}-
{\mathcal{R}}{\tilde{h}}\sum_{k=0}^{P-1}{{{b_k}}{\Gamma}^{(1)}_{m,k}}{\Big|}^2}}\right).
\end{align}

In the limit as $F$ approaches infinity, the maximum-likelihood estimates of symbols, $\underline{\hat{b}}_P$, are those that
maximize $f(\underline{y}_{\rm rec}^{(F\rightarrow\infty)}{|}\underline{b}_P)$, i.e.,
\begin{align} \label{eq40}
\underline{\hat{b}}_P&=\underset{\underline{b}_P}{\operatorname{\rm arg~max}}~\lim_{F\rightarrow\infty }
\int_{0}^{\infty}{{f}(\underline{y}_{\rm rec}^{(F)}|\underline{b}_P,\tilde{h})}f(\tilde{h})d\tilde{h}\nonumber\\
&=\underset{\underline{b}_P}{\operatorname{\rm arg~max}}~\lim_{F\rightarrow\infty }
\int_{-\infty}^{\infty}{{f}(\underline{y}_{\rm rec}^{(F)}|\underline{b}_P,X)}f(X)dX,
\end{align}
where $X=0.5\ln(\tilde{h})$ is the log-amplitude factor of fading with mean $\mu_X$ and variance $\sigma^2_X$. Further simplifications on Eq. \eqref{eq40} results into the following criteria as the MSD algorithm for the estimation of the transmitted symbols in a SISO UVLC systems with BPPM signaling;
\begin{align} \label{eq41}
&\underline{\hat{b}}_P=\underset{\underline{b}_P}{\operatorname{\rm arg~max}}\int_{-\infty}^{\infty}\exp{\Bigg(}-\frac{1}{2\sigma^2_X}(X-\mu_X)^2 -\frac{1}{2\sigma^2_{Z_{m}}}\nonumber\\
&{{\sum_{m=1}^{\infty}{{\Big|}{y_{{\rm rec},m}}\!-\!{\mathcal{R}}{e^{2X}}\sum_{k=0}^{P-1}{{\overline{b_k}}{\Gamma}^{(0)}_{m,k}}\!-\!{\mathcal{R}}{e^{2X}}\sum_{k=0}^{P-1}{{{b_k}}{\Gamma}^{(1)}_{m,k}}{\Big|}^2}}}{\Bigg)}dX.
\end{align}

One may notice that, due to the integration over the fading coefficient, the MSD algorithm in \eqref{eq41} results into an intractable form. Here, we propose a low-complexity detection method based on the generalized multiple-symbol detection (GMSD) \cite{chatzidiamantis2010generalized}. The GMSD algorithm is a two-step process 
which estimates  both  the  transmitted  sequence  and  the  fading  coefficient  using the 
observation window of the consecutive received symbols. Based on \eqref{eq39} and the Parseval's theorem, ${{\sum_{m=1}^{\infty}{|{Z_{m}}|^2}} }=\int_{-\infty}^{\infty}|Z(t)|^2dt$, the GMSD algorithm for the SISO UVLC systems with BPPM signaling can be expressed as;
\begin{align} \label{eq42}
&\underline{\hat{b}}_P= \underset{\underline{b}_P}{\operatorname{\rm arg~max}}~~~ \mathcal{G}(\underline{b}_P,\tilde{h}|\tilde{h}=\hat{\tilde{h}})
\end{align}
where the GMSD estimation function is obtained as;
\begin{align} \label{eq43}
\mathcal{G}(\underline{b}_P,\tilde{h})=&
-\sum_{k=0}^{P-1}{\sum_{k'=0}^{P-1}2{\mathcal{R}}{{\tilde{h}}^2{b_k}\overline{b_{k'}}x_{0,1}((k-k')T)}}\nonumber\\
&-\sum_{k=0}^{P-1}{\sum_{k'=0}^{P-1}{{\mathcal{R}}{\tilde{h}}^2\overline{b_k}~\overline{b_{k'}}x_{0,0}((k-k')T)}}\nonumber\\
& 
+\sum_{k=0}^{P-1}2{{\tilde{h}}\overline{b_k}r_{0}(kT)}+\sum_{k=0}^{P-1}2{{\tilde{h}}{b_k}r_{1}(kT)}\nonumber\\
&-\sum_{k=0}^{P-1}{\sum_{k'=0}^{P-1}{\mathcal{R}}{{\tilde{h}}^2{b_k}~{b_{k'}}x_{1,1}((k-k')T)}},
\end{align}
in which $r_{0}(t)=y_{\rm rec}\left(t\right)*{\Gamma}^{(0)}(-t)$, $r_{1}(t)=y_{\rm rec}\left(t\right)*{\Gamma}^{(1)}(-t)$, $x_{0,0}(t)={\Gamma}^{(0)}(t)*{\Gamma}^{(0)}(-t)$, $x_{1,1}(t)={\Gamma}^{(1)}(t)*{\Gamma}^{(1)}(-t)$, and $x_{0,1}(t)={\Gamma}^{(0)}(t)*{\Gamma}^{(1)}(-t)$. 
Moreover, the estimation of the fading coefficient, used in \eqref{eq42}, can be calculated as;
\begin{align} \label{eq44}
&\hat{\tilde{h}}= \underset{\tilde{h}}{\operatorname{\rm arg~max}}~~~ \mathcal{G}(\underline{b}_P,\tilde{h}).
\end{align}
By equating the derivative of $\mathcal{G}(\underline{b}_P,\tilde{h})$ in Eq. \eqref{eq43}, with respect to $\tilde{h}$, to zero, we can obtain the estimated value of the fading coefficient as Eq. \eqref{eq45}, in the top of the next page.
\begin{figure*}[t!]
\normalsize
\setcounter{equation}{44}
\begin{align}\label{eq45}
\hat{\tilde{h}}=\frac{ {\underset{k=0}{\overset{P-1}{\sum}}}~\!{\overline{b_k}r_{0}(kT)}+{\underset{k=0}{\overset{P-1}{\sum}}}{{b_k}r_{1}(kT)}}{{\underset{k=0}{\overset{P-1}{\sum}}}{\underset{k'=0}{\overset{P-1}{\sum}}}{2{\mathcal{R}}{{b_k}\overline{b_{k'}}x_{0,1}((k-k')T)}}+{\underset{k=0}{\overset{P-1}{\sum}}}{\underset{k'=0}{\overset{P-1}{\sum}}}{{{\mathcal{R}}\overline{b_k}~\overline{b_{k'}}x_{0,0}((k-k')T)}}+{\underset{k=0}{\overset{P-1}{\sum}}}{\underset{k'=0}{\overset{P-1}{\sum}}}{{\mathcal{R}}{{b_k}~{b_{k'}}x_{1,1}((k-k')T)}}}.
\end{align}
 \hrulefill
\end{figure*}

It is worth mentioning that for typical bit duration times, $\Gamma^{(0)}(t)$ and $\Gamma^{(1)}(t)$ are temporally spread over a limited number of bits. Consequently, the summations in Eq. \eqref{eq43} are in fact defined over a finite number of bits. Therefore, based on the slow fading nature of UVLC channels \cite{tang2013temporal,jamali2016statistical}, we can consider a constant fading coefficient for all of the involved consecutive bits in each summation of \eqref{eq43} for typical values of the bit duration time and detection window length $P$. Moreover, the GMSD algorithm which estimates the fading coefficient as \eqref{eq45} does not require any instantaneous and statistical channel state information and can be applied when the channel fading coefficient is modeled using any statistical distribution rather than lognormal distribution.

\section{Numerical Results and Discussion}
In this section, we present the numerical results and simulations for the performance of various UVLC system configurations in different water types, namely clear ocean, coastal, and turbid harbor waters with coefficients listed in Table \ref{T1} \cite{mobley1994light}. Both laser-based collimated and LED-based diffusive links are considered. Some of the important system parameters used for the channel MC simulation, including the transmitter and receiver specifications, are summarized in Table \ref{T2}.
\begin{table}[t]
\centering
\caption{Absorption, scattering, and extinction coefficients of different water types \cite{mobley1994light}.}
\label{T1}
 \begin{tabular}{c c c c}  
Water type		   & $a$ [\si{m^{-1}}]        &  $b$ [\si{m^{-1}}] &  $c$ [\si{m^{-1}}]  \\ \hline\hline
 Clear ocean& $0.114$ & $0.037$ & $0.151$   \\  
 Coastal water & $0.179$ & $0.219$ & $0.398$   \\
  Turbid harbor water & $0.366$ & $1.824$ & $2.190$  \\
\end{tabular}
\end{table}
\begin{table}[t]
\centering
\caption{Some of the important parameters used for the channel MC simulation.}
\label{T2}
  \begin{tabular}{||M{2.35in}||M{0.65in}||}
  \hline
 Coefficient & Value\\
   \hline \hline
Laser half divergence angle, $\theta_{div}$ & $0.75$ \si{mrad}\\ \hline
Laser beam waist radius, $W_r$ & $1$ \si{mm} \\ \hline
Semi-angle at half power of LED, $\theta_{1/2}$ & $15^{\rm o}$ \\ \hline
Source wavelength, $\lambda $ & $532$ \si{nm} \\ \hline 
Total number of transmitted photons for the channel MC simulation, $N_t$ & $10^8$ \\ \hline
Water refractive index, $n$ & $1.331$ \\ \hline 
   Receiver half angle FOV, $\theta_{\rm FOV}$ & ${40}^{\rm o}$ \\ \hline 
     MISO schemes aperture diameter, $D^{(\rm MISO)}_0$ & $20$ \si{cm} \\ \hline 
     Photon weight threshold at the receiver, $W_{th}$ & ${10}^{-6}$ \\ \hline 
    Center-to-center separation distance between the transmitters and between the receiving apertures, $l_0$ & $25$ \si{cm} \\ \hline
  \end{tabular}
  \end{table}
  
\begin{figure*}
\centering
\includegraphics[width=7in]{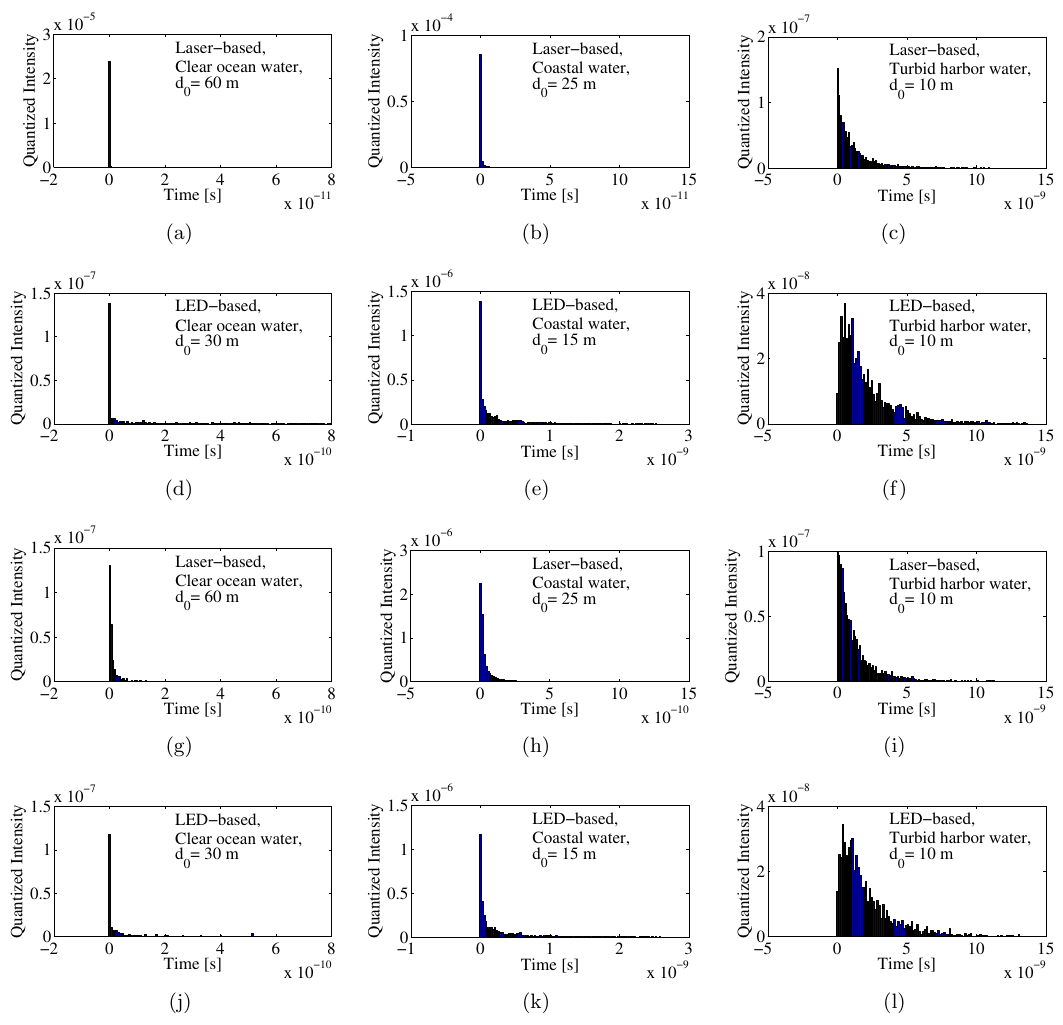}
\caption{FFIR of laser-based collimated and LED-based diffusive UVLC links with different water types, namely clear ocean, coastal, and turbid harbor waters. (a)-(f) FFIR of the direct link, $h_{0,11}(t)$; (g)-(l) FFIR of the indirect link, $h_{0,12}(t)$.}
\label{fig_FFIR_SISO}
\end{figure*}

For the channel MC simulation, we have generated $N_t=10^8$ photons at the transmitter side and recorded all attributes of detectable photons by circular receivers with aperture diameters of $D^{(\rm MISO)}_0/\sqrt{N}$, according to the discussed procedure in Section II-A. To illustrate the behavior of UVLC channels with various water types and link geometries, we consider a $1\times 3$ SIMO configuration. The transmitter is pointed to one of the receivers, namely the medial one which we name it as the first receiver. The direct link between the transmitter and first receiver has FFIR of $h_{0,11}(t)$. In addition to the aforementioned direct link, the other two receivers, which are located at the center-to-center distance of $l_0=25$ \si{cm} from the medial aperture, also receive the scattered photons of the transmitter because of the channel multiple scattering effect on the propagating photons. Due to the symmetrical geometry, both of the links from the transmitter to the top and down receivers have the same FFIR of $h_{0,12}(t)$.
Figs. \ref{fig_FFIR_SISO}(a)-(f) show the FFIR of the direct link, $h_{0,11}(t)$, for various water types and link geometries, while Figs. \ref{fig_FFIR_SISO}(g)-(l) depict $h_{0,12}(t)$ for the same scenarios. In order to obtain these plots, for each channel scenario, we have divided the difference of arrival times of the latest and earliest detected photons into $100$ bins. Then for each bin, we have calculated the sum of the weights of all detected photons in the corresponding time interval of that bin and normalized the summation to the total number of transmitted photons.
  
Comparing Figs. \ref{fig_FFIR_SISO}(a)-(c) shows that as the water turbidity increases the channel delay spread increases too. As it is obvious from these figures, the FFIR of laser-based clear ocean channels, even for long ranges as $d_0=60$ \si{m}, can properly be modeled by an ideal delta function. On the other hand, for turbid harbor waters, where the attenuation length $z_a=cd_0$ is large, even low-range (e.g., $10$ \si{m}) laser-based links significantly spread the received signal in temporal domain.
 Moreover, based on the details of the channel MC simulation \cite{cox2012simulation,cox2014simulating}, each photon may interact with water molecules and other particles after being propagated $\Delta s=-\frac{1}{c}\ln\zeta_s$ \si{m}, where $\zeta_s$ is a uniformly distributed RV in the interval $[0,1]$. And after each $i$th interaction the photon weight decreases as $W^{(i+1)}=(1-\frac{a}{c})W^{(i)}$. Therefore, as the channel extinction coefficient $c$ increases, due to the increased number of interactions, the channel attenuation becomes more severe and also the channel temporal spread grows because of the higher attenuation length. All these significantly deteriorate the turbid harbor channels' quality such that a $10$ \si{m} laser-based turbid harbor link has by far worse channel quality than a $25$ \si{m} laser-based coastal link and even a $60$ \si{m} clear ocean link.
  
The results of the direct link FFIR in Figs. \ref{fig_FFIR_SISO}(d)-(f) illustrate that LED-based transmission increases the channel diffusivity and therefore the channel loss and temporal dispersion. In other words, increasing the transmitter divergence angle turns the transmitted photons away from the direct path and hence, decreases the probability of these photons to be captured by the receiver leading to higher transmission loss. Moreover, the transmitted photons with larger zenith angles will probably experience more scattering and interactions with water particles to reach the receiver. This fact, in addition to increasing the channel attenuation (due to the photons' weight reduction), increases the channel delay spread for diffusive LED-based links. Therefore, as an example, a $30$ \si{m} LED-based clear ocean link has worse channel quality, i.e., higher attenuation and delay spread, than the laser-based counterpart even with the double link length. Comparing Figs. \ref{fig_FFIR_SISO}(a)-(f) shows that the aforementioned problem is less concerning for LED-based coastal water links and especially turbid harbor water links. In other words, when the channel extinction coefficient or better said the channel attenuation length is large, even when the propagating photons are transmitted with very small divergence angles, the large number of interactions of photons causes them to deviate from the direct path and also reach the receiver plane with abundant number of scattering and hence with high transmission time differences, i.e., relatively a similar behavior to LED-based transmission is observable. Therefore, as expected, switching from laser to LED-based transmission does not severely degrade the channel quality for turbid harbor water links.
  
  \begin{figure*}[t]
  \centering
  \includegraphics[trim=0cm 0.1cm 0cm 0cm,width=6in,clip]{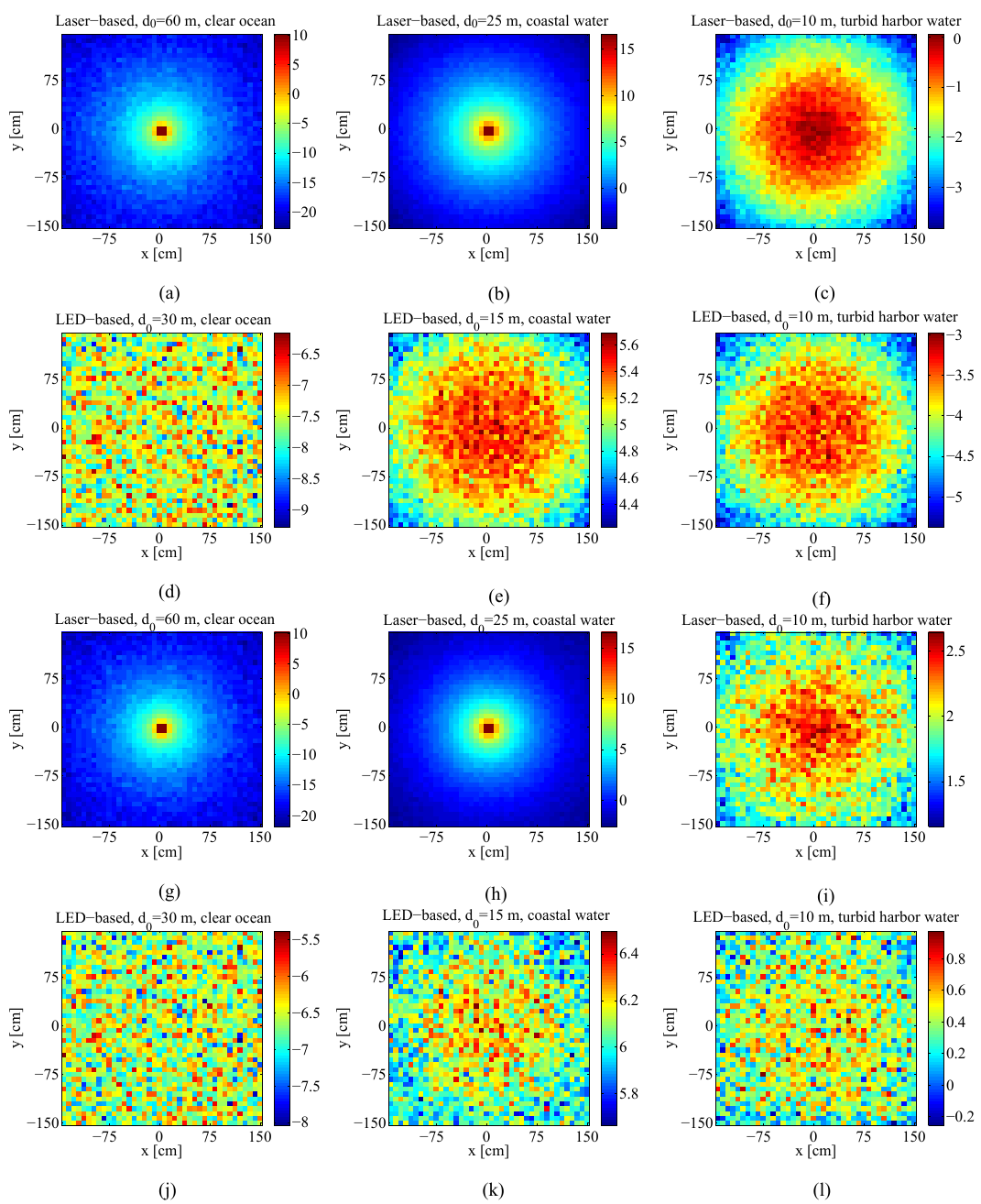}
  \caption{Spatial distribution of the illumination over a $3\times 3$ \si{m^2} square area of the receiver plane (with center in $(x,y)=(0,0)$) for the discussed six scenarios in Fig. 2. (a)-(f) Integration time $T_0=1$ \si{ns}; (g)-(l) integration time $=5T_0=5$ \si{ns}.}
  \label{fig_spatial_distribution}
  \vspace{-0.18in}
  \end{figure*}
  
Figs. \ref{fig_FFIR_SISO}(g)-(l) depict the indirect link FFIR, $h_{0,12}(t)$, for the same scenarios as Figs. \ref{fig_FFIR_SISO}(a)-(f). Note that the transmitter is pointed to the medial receiver and the second receiver is located in the center-to-center distance of $l_0=25$ \si{cm} from the medial one and detects those photons that arrive with multiple scattering, from the indirect path with FFIR of $h_{0,12}(t)$. Therefore, it is expected that $h_{0,12}(t)$ experience more attenuation and delay spread than $h_{0,11}(t)$. 
Comparing the results in Fig. \ref{fig_FFIR_SISO} shows that when the laser-based transmission is employed in clear ocean or coastal water, for which the channel attenuation length is relatively small and the transmitter beam focus approximately endures until the receiver, the channel spatial beam spread is negligible and hence, the main energy of the transmitted signal will be captured by the medial receiver. In such cases, $h_{0,12}(t)$ has very poor condition, i.e., higher loss and temporal dispersion, than $h_{0,11}(t)$.
On the other hand, when either the link or transmitter is highly diffusive, i.e., when turbid harbor channel is considered or LED is used at the transmitter, there is no any considerable difference between $h_{0,11}(t)$ and $h_{0,12}(t)$. In other words, the diffusive nature of the channel spatially spreads photons in the receiver plane and leads to an approximately uniform illumination over a larger area than collimated links.
 Therefore, the tracking and alignment problems for collimated laser-based links in clear ocean and coastal waters are by far more challenging and critical than diffusive LED-based links and also laser-based links in turbid harbor waters. Hence, although LED-based transmission increases both the channel attenuation and temporal dispersion, when a relatively low-speed mobile underwater optical communication is required, the diffusive links employment is inevitable to decrease the tracking and alignment difficulties. Such diffusive links can potentially be used for low-range mobile applications, e.g., between the mobile users and first relay of the proposed cellular relay-assisted CDMA-based UVLC network in \cite{jamali2016relay}. On the other hand, when we are interested to a fixed point-to-point underwater communication, e.g., between the fixed relay nodes of the mentioned network, it is preferred to employ collimated laser-based links to increase both the viable communication range and the transmission data rate.

In order to better portray the channel spatial spread, in Fig. \ref{fig_spatial_distribution} we illustrate the spatial distribution of the illumination over a $3\times 3$ \si{m^2} square area of the receiver plane (with center in $(x,y)=(0,0)$) for the discussed six scenarios in Fig. \ref{fig_FFIR_SISO}. To do so, we divide this area into $40\times 40$ pixels, each with a square area of $7.5\times 7.5$ \si{cm^2}. 
Then for each square pixel, we calculate the normalized intensity over time slots of $T_0=1$ \si{ns}, from the time of the first detected photon. The value of each $(i,j)$th pixel in Figs. \ref{fig_spatial_distribution}(a)-(f) is calculated as $10\log_{10}\left(\sum_{s_1\in\Lambda_{1,ij}}W_{s_1}/N_t\right)$, where $W_{s_1}$ is the weight of the $s_1$th detected photon, and $\Lambda_{1,ij}$ specifies the set of all detected photons within the $(i,j)$th spatial pixel and the first $T_0$ seconds time interval. For Figs. \ref{fig_spatial_distribution}(g)-(l) we extend the time window from $T_0$ to $5T_0=5$ \si{ns} and repeat the previous six observations to investigate the effect of integration time on the spatial illumination pattern.

Figs. \ref{fig_spatial_distribution}(a) and \ref{fig_spatial_distribution}(b) show that for laser-based clear ocean and coastal water links, with typical communication ranges, the beam spatial spread is negligible and the received energy is mainly confined around the beam center. Although this behavior improves the channel quality by decreasing both the channel loss and temporal dispersion, as discussed before, it increases the tracking and alignment difficulties for such collimated link geometries. Comparing these figures with Figs. \ref{fig_spatial_distribution}(g) and \ref{fig_spatial_distribution}(h) confirms that the channel temporal spread is also negligible for such collimated links. In other words, increasing the integration time does not considerably change the spatial pattern and the intensity value of each spatial pixel, i.e., the main energy of all spatial pixels are accumulated in a small time interval; within the first $T_0$ seconds.

Figs. \ref{fig_spatial_distribution}(c)-(f) demonstrate that for laser-based transmission in turbid harbor waters and also all LED-based links the channel considerably spreads the propagating signal in spatial domain, such that the transmitted signal, approximately, uniformly illuminates a large area of the receiver plane. As discussed before, such a behavior of diffusive channels, however deteriorates the link quality by increasing both the channel loss and delay spread, alleviates the tracking and alignment troubles and enables such diffusive links to support underwater users' mobility. For example, an acceptably uniform energy reception over a $3\times 3$ \si{m^2} area provides up to one second continuous communication for a low-speed underwater user with $3$ \si{m/s} mobility. Comparing these figures with Figs. \ref{fig_spatial_distribution}(i)-(l) verifies that the aforementioned diffusive links also significantly spread the received signal in temporal domain, such that increasing the integration time increases the normalized intensity of each spatial pixel, i.e., the detected energy is distributed over much larger time intervals than $T_0=1$ \si{ns}. Moreover, the above comparison indicates that as the integration time increases the illumination uniformity improves, i.e., the difference between the normalized intensity of the brightest and darkest pixels decreases. Therefore, it is reasonable to transmit with lower data rates through such diffusive links to decrease the channel loss and ISI, and also improve the uniformity of the received intensity over larger areas for mobile applications.

\begin{table*}[t]
  \centering
  \caption{The channel first $5$ loss coefficients of both the direct and indirect links with $T_b=1$ \si{ns} as well as RMS delay spread for different scenarios. Here, the aperture area of each receiver is assumed to be $20$ \si{cm}.}
  \label{T3}
   \begin{tabular}{M{1cm}||M{2cm}M{2cm}M{2.5cm}M{2cm}M{2cm}M{2.5cm}}  
Channel Condition & Laser-based $d_0=60$ \si{m} ~ clear ocean &  Laser-based $d_0=25$ \si{m} coastal water & Laser-based $d_0=10$ \si{m} ~~~ turbid harbor water & LED-based $d_0=30$ \si{m} ~ clear ocean & LED-based $d_0=15$ \si{m} coastal water & LED-based ~~ $d_0=10$ \si{m} ~~~~~ turbid harbor water \\ \hline\hline
  $\rho_{0,11}$& $2.4889\times 10^{-5}$& $9.8355\times 10^{-5}$&$4.1449\times 10^{-7}$&
  $2.0403\times 10^{-7}$& $2.8139\times 10^{-6}$&$2.5818\times 10^{-7}$\\  
  $\rho_{1,11}$&$6.3993\times 10^{-9}$ & $3.5116\times 10^{-7}$&$3.8509\times 10^{-7}$&
  $2.6305\times 10^{-8}$&$7.6994\times 10^{-7}$&$5.7212\times 10^{-7}$ \\
  $\rho_{2,11}$&$3.2052\times 10^{-10}$ &$2.0933\times 10^{-8}$&$1.5219\times 10^{-7}$&
  $6.1243\times 10^{-9}$& $1.5789\times 10^{-7}$& $3.6744\times 10^{-7}$\\
  $\rho_{3,11}$&$8.8050\times 10^{-11}$ &$7.0432\times 10^{-9}$&$7.1750\times 10^{-8}$&
  $4.9795\times 10^{-9}$& $5.9260\times 10^{-8}$& $2.0638\times 10^{-7}$\\
  $\rho_{4,11}$&$4.8756\times 10^{-11}$ &$3.0451\times 10^{-9}$&$3.8263\times 10^{-8}$&
  $2.0023\times 10^{-9}$& $3.0219\times 10^{-8}$& $1.2395\times 10^{-7}$\\
  \hline
  $\rho_{0,12}$&$2.7642\times 10^{-7}$&$5.8637\times 10^{-6}$&$3.6722\times 10^{-7}$&
  $1.9169\times 10^{-7}$& $2.7827\times 10^{-6}$& $2.4466\times 10^{-7}$ \\  
  $\rho_{1,12}$&$6.5210\times 10^{-9}$&$3.8624\times 10^{-7}$&$4.1149\times 10^{-7}$&
  $2.7404\times 10^{-8}$& $7.5837\times 10^{-7}$& $5.8861\times 10^{-7}$ \\
  $\rho_{2,12}$&$1.8550\times 10^{-10}$&$2.6422\times 10^{-8}$&$1.5989\times 10^{-7}$& 
  $9.6763\times 10^{-9}$& $1.5030\times 10^{-7}$&$3.7856\times 10^{-7}$ \\
  $\rho_{3,12}$&$3.9447\times 10^{-11}$&$7.4601\times 10^{-9}$&$7.1180\times 10^{-8}$&
  $2.7246\times 10^{-9}$& $6.7100\times 10^{-8}$& $2.0778\times 10^{-7}$ \\
  $\rho_{4,12}$&$1.0497\times 10^{-10}$&$3.4574\times 10^{-9}$&$3.9987\times 10^{-8}$&
  $1.0912\times 10^{-9}$& $2.3123\times 10^{-8}$& $1.1873\times 10^{-7}$ \\
  \hline
    $\tau_{{\rm RMS},11}$&$7.4\times 10^{-5}$ \si{ns}&$5.3\times 10^{-4}$ \si{ns}&$1.0413$ \si{ns}&
    $0.0834$ \si{ns}& $0.1139$ \si{ns}& $1.6485$ \si{ns}\\  
    \hline
        $\tau_{{\rm RMS},12}$&$0.0116$ \si{ns}&$0.0190$ \si{ns}&$ 1.1339$ \si{ns}&
        $0.2236$ \si{ns}& $0.1899$ \si{ns}& $1.5965$ \si{ns} \\  
  \end{tabular}
  \end{table*}
  \begin{table}[t]
      \centering
    \caption{Some of the important parameters used for the receiver noise characterization \cite{jaruwatanadilok2008underwater,giles2005underwater}.}
    \label{T4}
      \begin{tabular}{||M{2.2in}||M{0.8in}||}
        \hline
     Coefficient & Value\\
       \hline \hline
       Quantum efficiency, $\eta $ & $0.8$ \\ \hline 
                Optical filter bandwidth, $\triangle \lambda $ & $10$ \si{nm} \\ \hline 
                Optical filter transmissivity, $T_F$ & $0.8$ \\ \hline 
                Equivalent temperature, $T_e$ & $290$ \si{K} \\ \hline 
                Load resistance, $R_L$ & $100~\Omega $ \\ \hline 
                Dark current, $I_{dc}$ & $1.226\times {10}^{-9}$ \si{A}\\ \hline
                Downwelling irradiance, $E$ & $1440$ \si{W/m^2} \\ \hline
                Underwater reflectance of the downwelling irradiance, $R_d$ & $0.0125$ \\ \hline
                Viewing angle, $\phi_v$ & $90^{\rm o}$ (horizontal) \\ \hline
                Describing factor of the directional dependence of the underwater radiance, $L_{fac}$ & $2.9$ \\ \hline
                Water depth, $D_w$ & $30$ \si{m}\\ \hline
      \end{tabular}
      \end{table}

In order to more quantitatively discuss on the channel loss for various scenarios, we consider a rectangular pulse $p(t)$ with unit amplitude in the interval $[0,T_b]$ at the transmitter side, where $T_b$ is the bit duration time which for OOK modulation relates to the transmission data rate, $R_b$, as $T_b=1/R_b$. Then we numerically measure the received energy, after the pulse propagation through the channel with FFIR of $h_{0,ij}(t)$, over each $k$th time slot and normalize it to the transmitted pulse energy to obtain the channel loss coefficient of the $k$th time slot as;
\begin{align}\label{loss}
\rho_{k,ij}\!=\!\frac{\int_{kT_b}^{(k+1)T_b}p(t)*h_{0,ij}(t)~dt}{\int_{0}^{T_b}p(t)~dt}\!=\!\frac{1}{T_b}\int_{kT_b}^{(k+1)T_b}\!\!\!\!\!\!\!\!\!\!\!p(t)*h_{0,ij}(t)~dt.
\end{align}
Furthermore, we have calculated the root mean square (RMS) delay spread for different channel scenarios using;
\begin{align}\label{rms}
\tau_{{\rm RMS},ij}=\sqrt{\frac{\int_{-\infty}^{\infty}(t-\tau_{0,ij})^2h_{0,ij}^2(t)dt}{\int_{-\infty}^{\infty}h_{0,ij}^2(t)dt}},
\end{align}
where the mean delay time is given by;
\begin{align}\label{tau_0}
\tau_{0,ij}={\frac{\int_{-\infty}^{\infty}t.h_{0,ij}^2(t)dt}{\int_{-\infty}^{\infty}h_{0,ij}^2(t)dt}}.
\end{align}
Table \ref{T3} shows the first $5$ loss coefficients of both the direct and indirect links with $T_b=1$ \si{ns} as well as the RMS delay spread for different channel conditions, which definitely confirm our previous discussions. 

After the above comprehensive channel study, we turn our attention to the performance of MIMO UVLC systems with BPPM signaling and EGC. For characterizing the receiver noise components we assume typical parameter values listed in Table \ref{T4}. We further consider the receiver electronic bandwidth as $B=1/T$ \cite{jazayerifar2006atmospheric}. With respect to these parameters and the detailed expressions for the characterization of different noise components in \cite{jaruwatanadilok2008underwater} and \cite{giles2005underwater}, we found the received background power at $30$ \si{m} depth as $P_{BG}\approx 25.57\times 10^{-9}\times \exp(-30K_d)$ \si{W}, where $K_d$ is the diffuse attenuation factor \cite{mobley1994light} which for different water types ranges from around $0.04$ to $4$ \si{m^{-1}} \cite{lee2005diffuse}, and as the water turbidity increases this coefficient increases too \cite{mobley1994light}. Therefore, it can easily be justified that the background noise has a negligible effect on the performance of our system. 

Fig. 4 shows the BER performance of a UVLC link with $\sigma^2_X=0.16$ and $R_b=1$ \si{Gbps} in different water types for both LED-based diffusive and laser-based collimated links. As expected, the system performance considerably degrades for links with higher water turbidity. For example, because of its remarkably higher attenuation and scattering, a $10$ \si{m} laser-based harbor water link has by far worse performance than a clear ocean link with even six times longer link range. It is worth to be mentioned that for a $10$ \si{m} LED-based harbor water link with $1$ \si{Gbps} data rate, the interference between the received symbols is so much that even increasing the transmitted power cannot improve the system performance. Moreover, an excellent match between the analytical and photon-counting methods confirms the negligibility of shot noise effect while the good match with numerical simulations corroborates the accuracy of our derived analytical expressions.
\begin{figure}\label{fig4}
\centering
\includegraphics[width=3.4in]{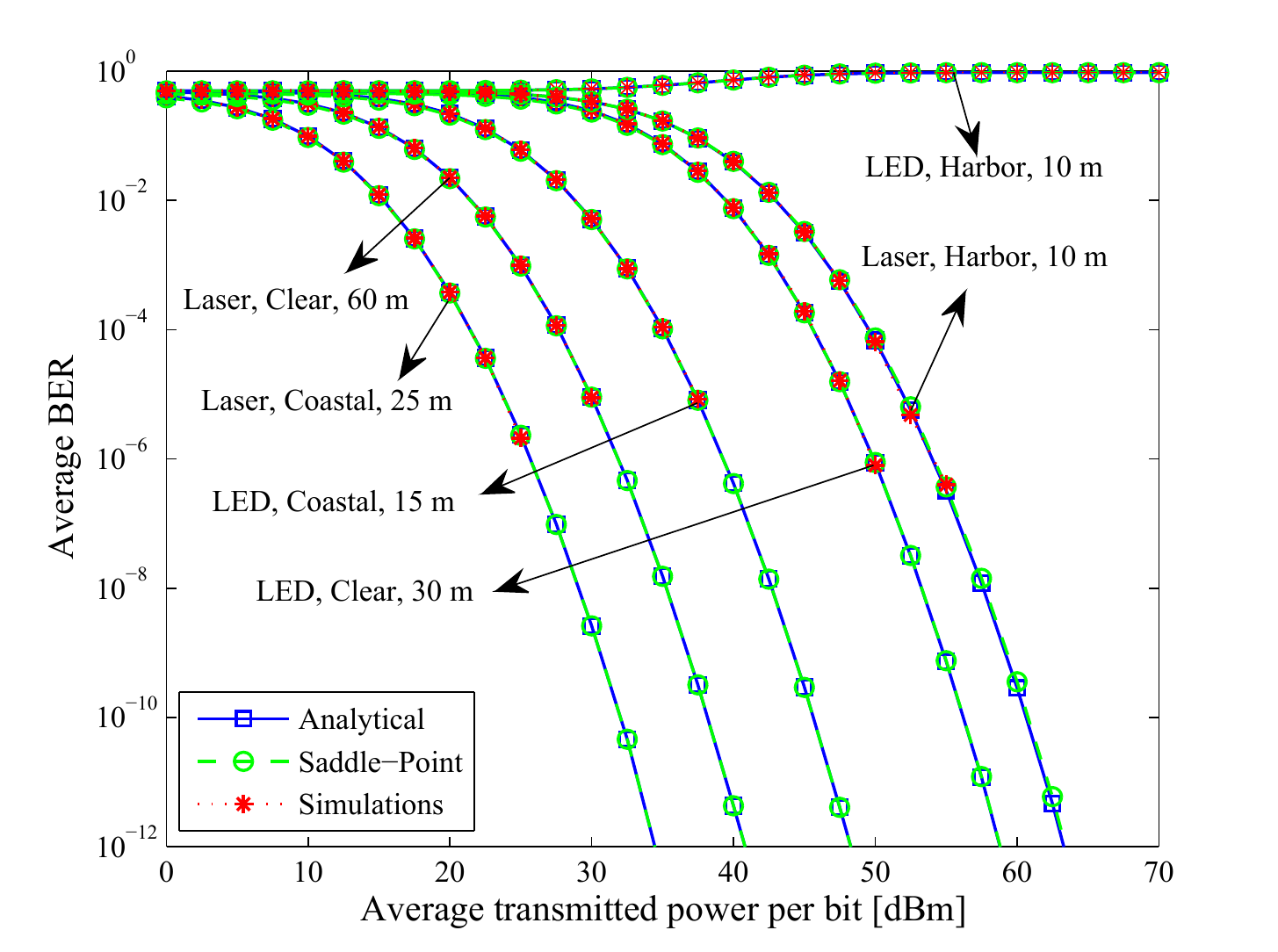}
\caption{Analytical, saddle-point, and simulation results for the exact BER of a SISO UVLC link with $\sigma^2_X=0.16$, $R_b=1$ \si{Gbps}, and different channel conditions.}
\label{}
\vspace{-0.4cm}
\end{figure}

Fig. 5 illustrates the BER curves for different configurations, including SISO, $2\times 1$ MISO, $3\times 1$ MISO, $1\times 2$ SIMO, $1\times 3$ SIMO, and $2\times 2$ MIMO, in a $6$ \si{m} LED-based harbor water link with $R_b=100$ \si{Mbps} and $\sigma^2_X=0.16$. As it can be seen, transmit diversity considerably improves the system performance, e.g., a $3\times 1$ MISO transmission provides about $8$ \si{dB} performance gain at the BER of $10^{-9}$ compared to SISO scheme. However, receiver diversity configurations have a slightly lower performance than transmit diversity, mainly because we have divided the total receiving aperture area by the number of receivers and also each receiver introduces a new thermal noise component. In addition, it is evident that increasing the number of independent branches, i.e., $M\times N$, increases the BER curves' slopes and gives higher diversity gains. Furthermore, excellent matches between the results proves the accuracy of our derived analytical and photon-counting BER expressions.
\begin{figure}\label{fig5}
\centering
\includegraphics[width=3.4in]{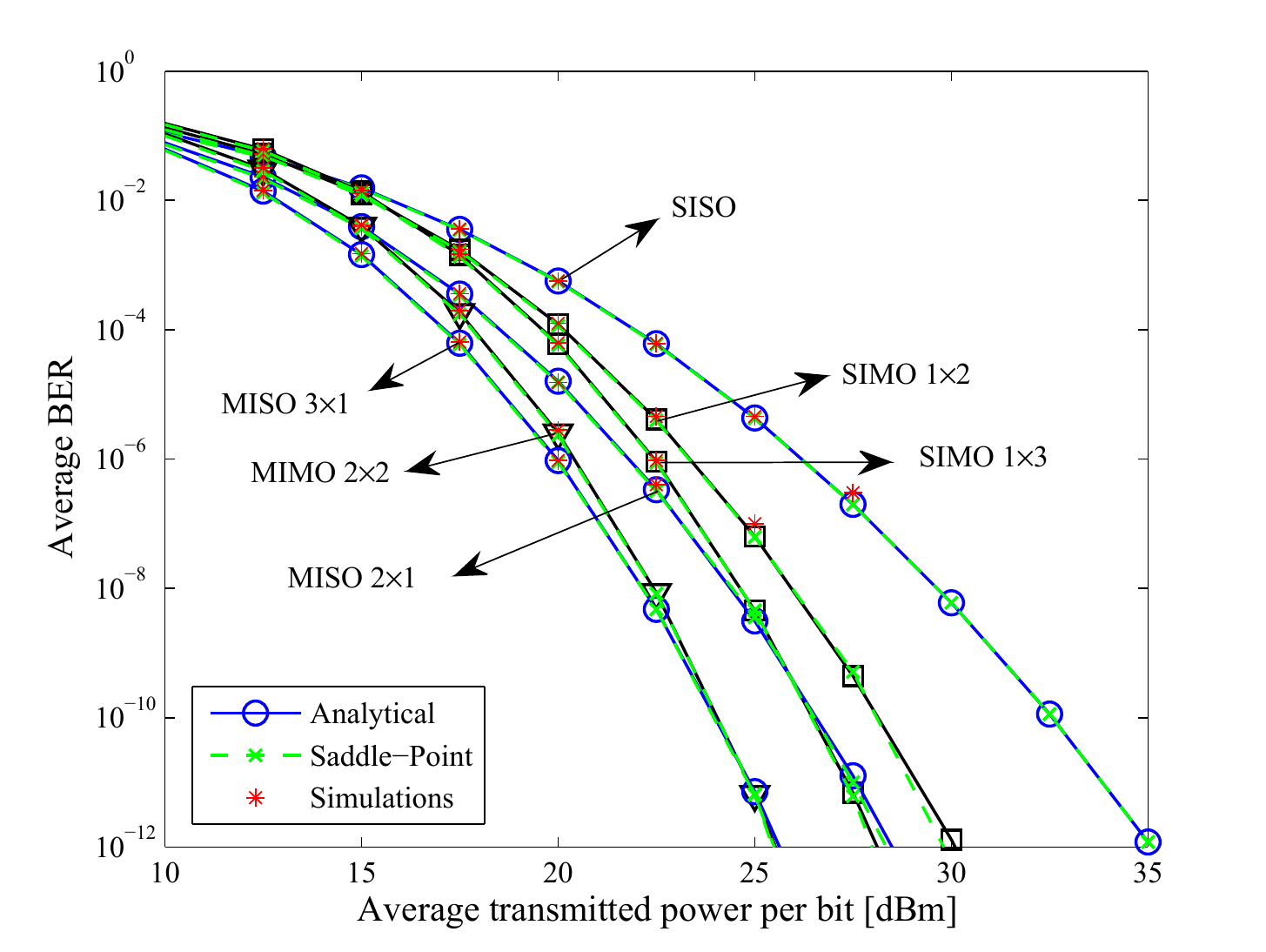}
\caption{BER performance of different configurations in a $6$ \si{m} LED-based harbor water link with $R_b=100$ \si{Mbps} and $\sigma^2_X=0.16$.}
\label{}
\vspace{-0.4cm}
\end{figure}

Fig. 6 compares the results obtained from ($M\times N$)-dimensional series of Gauss-Hermite quadrature formula (GHQF) as well as the approximated one-dimensional integral of Eqs. \eqref{SecIII-B-7}-\eqref{sigma2} with the exact ($M\times N$)-dimensional integrals for the BER evaluation of a UVLC system implemented in a $60$ \si{m} laser-based clear ocean link with $R_b=5$ \si{Gbps} and $\sigma^2_X=0.16$. As it is observable, while the approximated one-dimensional integration slightly overestimates the system BER, GHQF, using only $V_{ij}=30$ points, can accurately predict the exact BER of the system; this clearly demonstrates the advantage of GHQF in effective calculation of the system BER. Moreover, this figure shows that for UVLC channels with negligible spatial spread, SIMO schemes may even work worse than SISO transmission. In particular, based on Fig.3(a), the channel spatial spread for a $60$ \si{m} clear ocean link is negligible and hence the other off-axis receivers do not receive any considerable energy, as it is obvious by comparing Figs. 2(a) and 2(g). Therefore, dividing the total receiving aperture area by the number of receivers may even decrease the received energy while each receiver itself adds an independent thermal noise component. These altogether may lead to a worse performance than SISO transmission.
\begin{figure}\label{fig6}
\centering
\includegraphics[width=3.4in]{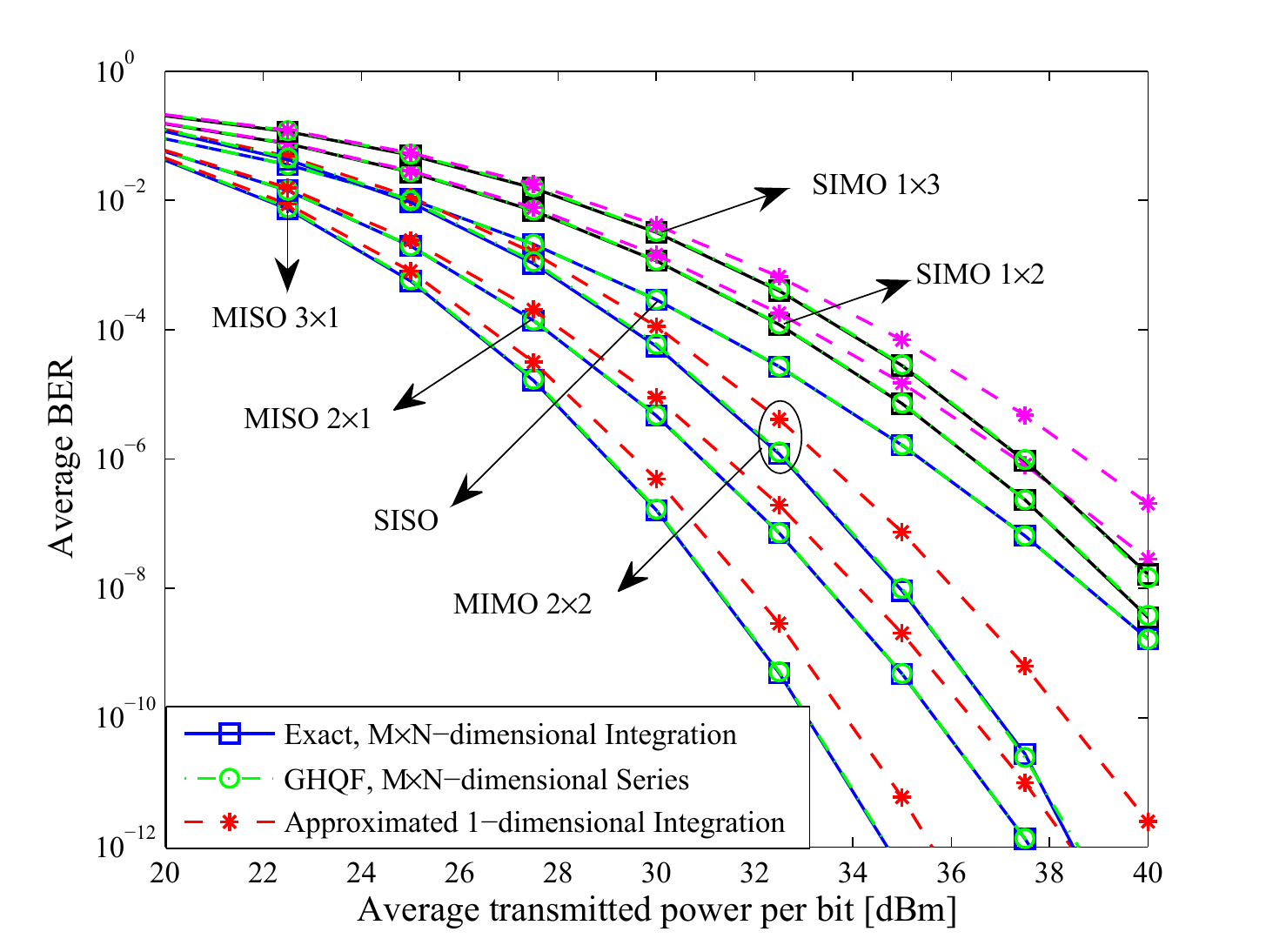}
\caption{Comparison between the results of different methods in predicting the performance of various configurations established in a $60$ \si{m} laser-based clear ocean link with $R_b=5$ \si{Gbps} and $\sigma^2_X=0.16$.}
\label{}
\vspace{-0.4cm}
\end{figure}

The effect of channel delay spread on the performance of UVLC systems is investigated in Fig. 7. In this figure, the BER performance of an LED-based SISO UVLC link with $\sigma^2_X=0.16$ is shown for two different scenarios, i.e., a $6$ \si{m} harbor water link and a $15$ \si{m} coastal water link, with various transmission data rates. As it can be seen, increasing the transmission data rate significantly deteriorates the system performance, due to the considerably increased ISI, especially for channels with larger delay spread. For example, while a $6$ \si{m} LED-based harbor water link has an excellent performance for $R_b=100$ \si{Mbps}, increasing the transmission data rate to $500$ \si{Mbps} demands excessively large average transmitted powers and makes the reliable communication infeasible.
 This behavior necessitates effective detection algorithms to improve the system performance in highly diffusive channels and make the reliable underwater communication possible for realistic transmitted powers per bit; the motivation behind the (G)MSD algorithm(s) derivation in this paper.
\begin{figure}\label{fig7}
\centering
\includegraphics[width=3.4in]{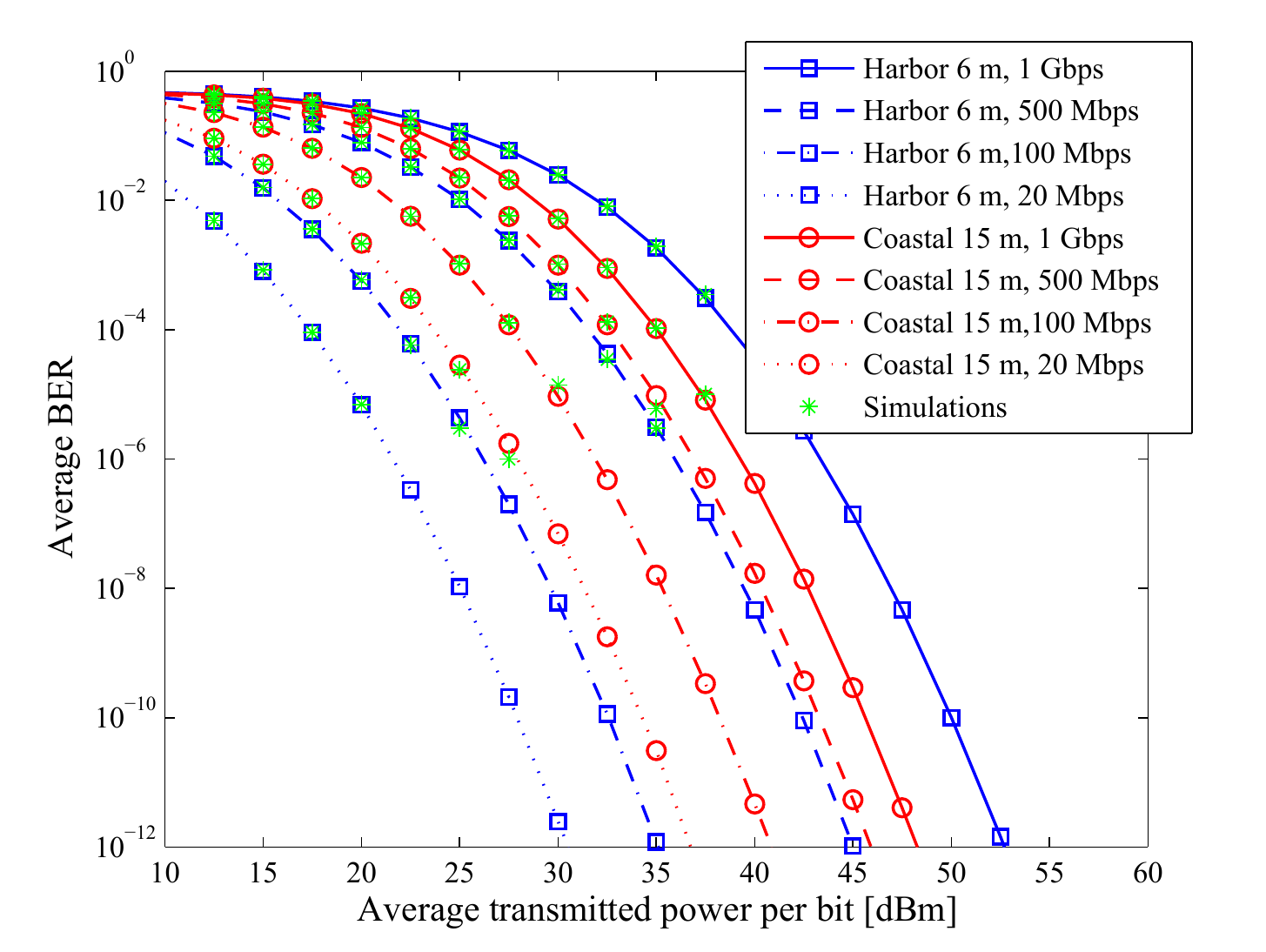}
\caption{Comparison between the performance results of two different scenarios, i.e., a $6$ \si{m} LED-based harbor water link and a $15$ \si{m} LED-based coastal water link, both with $\sigma^2_X=0.16$, for different transmission data rates.}
\label{}
\vspace{-0.4cm}
\end{figure}

In Fig. 8, we target illustrating the beneficial application of multiple-symbol detection, particularly the GMSD algorithm, in improving the performance of UVLC systems with high interference among the received symbols. For this purpose, we consider a $10$ \si{m} laser-based harbor water link with $\sigma^2_X=0.04$ and $R_b=500$ \si{Mbps}, and evaluate the performance of GMSD algorithm for the detection window lengths of $P=2$, $4$, and $10$. More specifically, for each block of consecutive bits, we first estimate the fading coefficient using Eq. \eqref{eq45}, and then apply this value in Eqs. \eqref{eq42} and \eqref{eq43} to detect the transmitted bits. As it is shown, using GMSD algorithm instead of symbol-by-symbol detection (SBSD) provides a remarkable performance improvement and makes the reliable underwater communications, for the considered link conditions, possible with realistic average transmitted powers per bits, which itself is equivalent to extension of the viable communication range. In particular, simultaneously detecting just two consecutive bits, i.e., $P=2$, improves the system BER by $7$ \si{dB} in comparison with SBSD at the BER of $10^{-6}$.
\begin{figure}\label{fig8}
\centering
\includegraphics[width=3.4in]{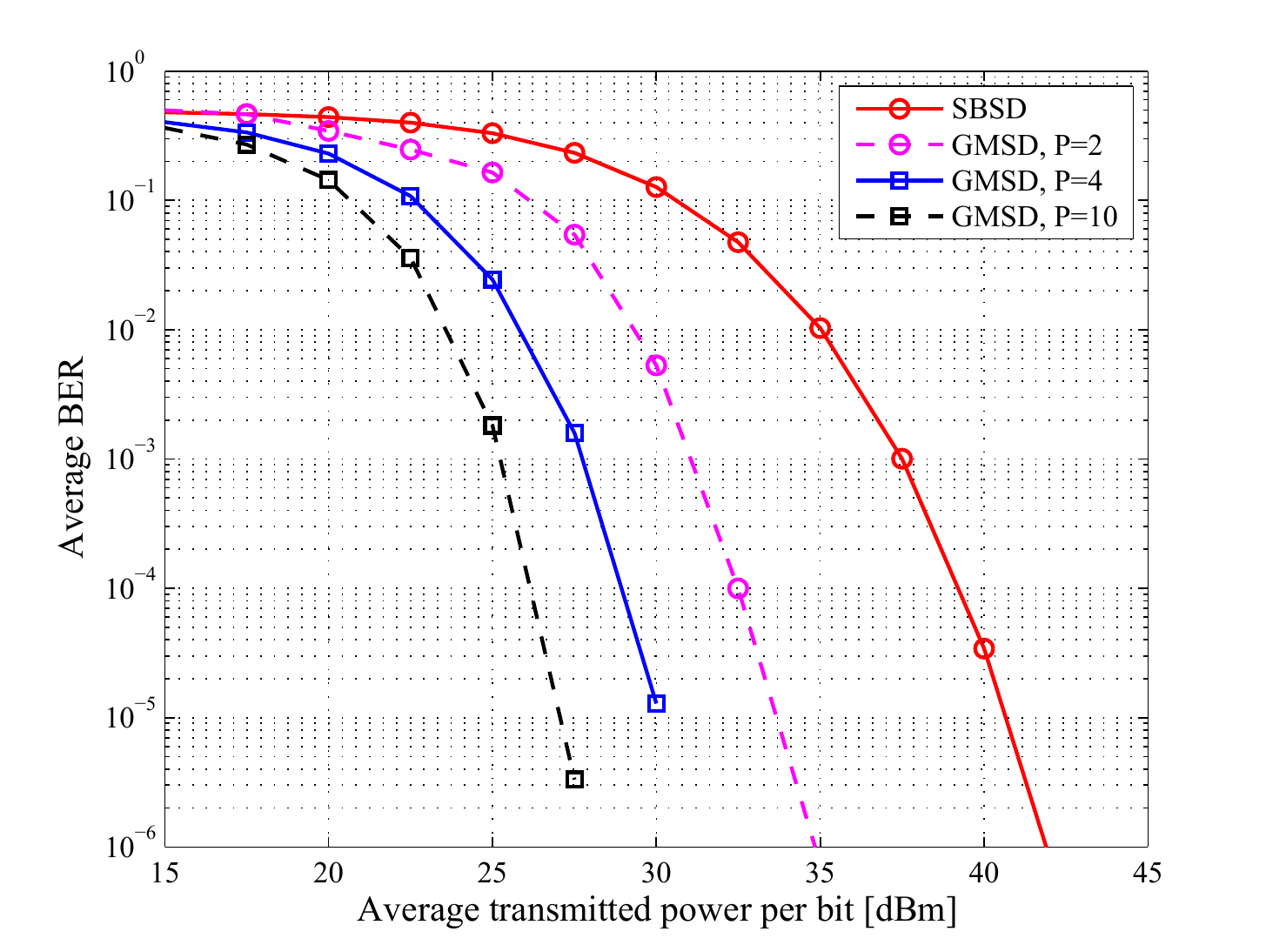}
\caption{GMSD algorithm results for the BER of a $10$ \si{m} laser-based harbor water link with $R_b=500$ \si{Mbps}, and different detection window lengths.}
\label{}
\vspace{-0.4cm}
\end{figure}
\section{Conclusion}
In this paper, we presented a comprehensive study over the channel characteristics, performance, and effective design of underwater visible light communications for both collimated laser-based and diffusive LED-based links with proper consideration of all of the channel degrading effects, including absorption, scattering, and turbulence-induced fading. We investigated the channel temporal and spatial spread to better observe its behavior and design our system. In order to tremendously improve the BER performance of UVLC systems and therefore increase their viable communication ranges, we applied MIMO technique to mitigate fading effects and multiple-symbol detection to alleviate ISI deteriorations. In addition to closed-form analytical expressions for the BER of MIMO UVLC systems with BPPM signaling, we also applied saddle-point approximation, which is based on photon-counting method, to take into account the shot noise effects. Excellent matches between our analytical results and simulations confirmed the accuracy of our derived expressions throughout the paper. Furthermore, a remarkable performance observed using both MIMO technique and (G)MSD algorithm(s). In particular, a $3\times 1$ MISO transmission provided about $8$ \si{dB} performance gain at the BER of $10^{-9}$ compared to SISO scheme. Also the GMSD algorithm with the detection window length of $P=2$, improved the system BER by $7$ \si{dB} in comparison with symbol-by-symbol detection at the BER of $10^{-6}$.


\begin{thebibliography}{10}
\baselineskip 12pt
\providecommand{\url}[1]{#1}
\csname url@samestyle\endcsname
\providecommand{\newblock}{\relax}
\providecommand{\bibinfo}[2]{#2}
\providecommand{\BIBentrySTDinterwordspacing}{\spaceskip=0pt\relax}
\providecommand{\BIBentryALTinterwordstretchfactor}{4}
\providecommand{\BIBentryALTinterwordspacing}{\spaceskip=\fontdimen2\font plus
\BIBentryALTinterwordstretchfactor\fontdimen3\font minus
  \fontdimen4\font\relax}
\providecommand{\BIBforeignlanguage}[2]{{%
\expandafter\ifx\csname l@#1\endcsname\relax
\typeout{** WARNING: IEEEtran.bst: No hyphenation pattern has been}%
\typeout{** loaded for the language `#1'. Using the pattern for}%
\typeout{** the default language instead.}%
\else
\language=\csname l@#1\endcsname
\fi
#2}}
\providecommand{\BIBdecl}{\relax}
\BIBdecl

\bibitem{tang2014impulse}
S.~Tang, Y.~Dong, and X.~Zhang, ``Impulse response modeling for underwater
  wireless optical communication links,'' \emph{IEEE Trans. Commun.}, vol.~62,
  no.~1, pp. 226--234, 2014.

\bibitem{hanson2008high}
F.~Hanson and S.~Radic, ``High bandwidth underwater optical communication,''
  \emph{Applied optics}, vol.~47, no.~2, pp. 277--283, 2008.

\bibitem{mobley1994light}
C.~D. Mobley, \emph{Light and water: Radiative transfer in natural
  waters}.\hskip 1em plus 0.5em minus 0.4em\relax Academic press, 1994.

\bibitem{petzold1972volume}
T.~J. Petzold, ``Volume scattering functions for selected ocean waters,'' SIO
  Ref. 72-78, Scripps Institution of Oceanography Visibility Laboratory, San
  Diego, CA, Tech. Rep., Oct. 1972.

\bibitem{jaruwatanadilok2008underwater}
S.~Jaruwatanadilok, ``Underwater wireless optical communication channel
  modeling and performance evaluation using vector radiative transfer theory,''
  \emph{IEEE J. Select. Areas Commun.}, vol.~26, no.~9, pp. 1620--1627, 2008.

\bibitem{zhang2016impulse}
H.~Zhang and Y.~Dong, ``Impulse response modeling for general underwater
  wireless optical {MIMO} links,'' \emph{IEEE Commun. Mag.}, vol.~54, no.~2,
  pp. 56--61, 2016.

\bibitem{akhoundi2015cellular}
F.~Akhoundi, J.~A. Salehi, and A.~Tashakori, ``Cellular underwater wireless
  optical {CDMA} network: Performance analysis and implementation concepts,''
  \emph{IEEE Trans. Commun.}, vol.~63, no.~3, pp. 882--891, 2015.

\bibitem{akhoundi2016cellular}
F.~Akhoundi, M.~V. Jamali, N.~Banihassan, H.~Beyranvand, A.~Minoofar, and J.~A.
  Salehi, ``Cellular underwater wireless optical {CDMA} network: Potentials and
  challenges,'' \emph{IEEE Access}, vol.~4, pp. 4254--4268, 2016.

\bibitem{jamali2016relay}
M.~V. Jamali, F.~Akhoundi, and J.~A. Salehi, ``Performance characterization of
  relay-assisted wireless optical {CDMA} networks in turbulent underwater
  channel,'' \emph{IEEE Trans. Wireless Commun.}, vol.~15, no.~6, pp.
  4104--4116, 2016.

\bibitem{jamali2016performance}
M.~V. Jamali, A.~Chizari, and J.~A. Salehi, ``Performance analysis of multi-hop
  underwater wireless optical communication systems,'' \emph{IEEE Photonics
  Technology Letters}, vol.~29, no.~5, pp. 462--465, 2017.

\bibitem{nikishov2000spectrum}
{V.V. Nikishov} and {V.I. Nikishov}, ``Spectrum of turbulent fluctuations of
  the sea-water refraction index,'' \emph{Int. J. Fluid Mech. Research},
  vol.~27, no.~1, pp. 82--98, 2000.

\bibitem{korotkova2012light}
O.~Korotkova, N.~Farwell, and E.~Shchepakina, ``Light scintillation in oceanic
  turbulence,'' \emph{Waves in Random and Complex Media}, vol.~22, no.~2, pp.
  260--266, 2012.

\bibitem{ata2014scintillations}
Y.~Ata and Y.~Baykal, ``Scintillations of optical plane and spherical waves in
  underwater turbulence,'' \emph{JOSA A}, vol.~31, no.~7, pp. 1552--1556, 2014.

\bibitem{gerccekciouglu2014bit}
H.~Ger{\c{c}}ekcio{\u{g}}lu, ``Bit error rate of focused {Gaussian} beams in
  weak oceanic turbulence,'' \emph{JOSA A}, vol.~31, no.~9, pp. 1963--1968,
  2014.

\bibitem{bohren2008absorption}
C.~F. Bohren and D.~R. Huffman, \emph{Absorption and scattering of light by
  small particles}.\hskip 1em plus 0.5em minus 0.4em\relax John Wiley \& Sons,
  2008.

\bibitem{cox2012simulation}
W.~C. Cox~Jr, \emph{Simulation, modeling, and design of underwater optical
  communication systems}.\hskip 1em plus 0.5em minus 0.4em\relax North Carolina
  State University, 2012.

\bibitem{cox2014simulating}
W.~Cox and J.~Muth, ``Simulating channel losses in an underwater optical
  communication system,'' \emph{JOSA A}, vol.~31, no.~5, pp. 920--934, 2014.

\bibitem{leathers2004monte}
R.~A. Leathers, T.~V. Downes, C.~O. Davis, and C.~D. Mobley, ``{Monte Carlo}
  radiative transfer simulations for ocean optics: a practical guide,'' DTIC
  Document, Tech. Rep., 2004.

\bibitem{tang2013temporal}
S.~Tang, X.~Zhang, and Y.~Dong, ``Temporal statistics of irradiance in moving
  turbulent ocean,'' in \emph{OCEANS-Bergen, 2013 MTS/IEEE}.\hskip 1em plus
  0.5em minus 0.4em\relax IEEE, 2013, pp. 1--4.

\bibitem{navidpour2007ber}
S.~M. Navidpour, M.~Uysal, and M.~Kavehrad, ``{BER} performance of free-space
  optical transmission with spatial diversity,'' \emph{IEEE Trans. Wireless
  Commun.}, vol.~6, no.~8, pp. 2813--2819, 2007.

\bibitem{andrews2001laser}
L.~C. Andrews, R.~L. Phillips, and C.~Y. Hopen, \emph{Laser beam scintillation
  with applications}.\hskip 1em plus 0.5em minus 0.4em\relax SPIE press, 2001.

\bibitem{zhu2002free}
X.~Zhu and J.~M. Kahn, ``Free-space optical communication through atmospheric
  turbulence channels,'' \emph{IEEE Trans. Commun.}, vol.~50, no.~8, pp.
  1293--1300, 2002.

\bibitem{karimi2009ber}
M.~Karimi and M.~Nasiri-Kenari, ``{BER} analysis of cooperative systems in
  free-space optical networks,'' \emph{J. Lightw. Technol.}, vol.~27, no.~24,
  pp. 5639--5647, 2009.

\bibitem{jamali2016statistical}
{M. V. Jamali et al.}, ``Statistical distribution of intensity fluctuations for
  underwater wireless optical channels in the presence of air bubbles,'' in
  \emph{4th Iran Workshop on Commun. and Inf. Theory (IWCIT)}.\hskip 1em plus
  0.5em minus 0.4em\relax IEEE, 2016, pp. 1--6.

\bibitem{yi2015underwater}
X.~Yi, Z.~Li, and Z.~Liu, ``Underwater optical communication performance for
  laser beam propagation through weak oceanic turbulence,'' \emph{Applied
  Optics}, vol.~54, no.~6, pp. 1273--1278, 2015.

\bibitem{lee2004part}
E.~J. Lee and V.~W. Chan, ``Part 1: Optical communication over the clear
  turbulent atmospheric channel using diversity,'' \emph{IEEE J. Select Areas
  Commun.}, vol.~22, no.~9, pp. 1896--1906, 2004.

\bibitem{jamali2015ber}
M.~V. Jamali and J.~A. Salehi, ``On the {BER} of multiple-input multiple-output
  underwater wireless optical communication systems,'' in \emph{4th
  International Workshop on Optical Wireless Communications (IWOW)}.\hskip 1em
  plus 0.5em minus 0.4em\relax IEEE, 2015, pp. 26--30.

\bibitem{jazayerifar2006atmospheric}
M.~Jazayerifar and J.~A. Salehi, ``Atmospheric optical {CDMA} communication
  systems via optical orthogonal codes,'' \emph{IEEE Trans. Commun.}, vol.~54,
  no.~9, pp. 1614--1623, 2006.

\bibitem{giles2005underwater}
J.~W. Giles and I.~N. Bankman, ``Underwater optical communications systems.
  part 2: basic design considerations,'' in \emph{Military Communications
  Conference, 2005. MILCOM 2005. IEEE}.\hskip 1em plus 0.5em minus 0.4em\relax
  IEEE, 2005, pp. 1700--1705.

\bibitem{einarsson2008principles}
G.~Einarsson, \emph{Principles of Lightwave Communications}.\hskip 1em plus
  0.5em minus 0.4em\relax New York: Wiley, 1996.

\bibitem{abramowitz1970handbook}
M.~Abramowitz and I.~A. Stegun, \emph{Handbook of mathematical functions: with
  formulas, graphs, and mathematical tables}.\hskip 1em plus 0.5em minus
  0.4em\relax Dover, New York, 1970.

\bibitem{jamali2015performanceMIMO}
M.~V. Jamali, J.~A. Salehi, and F.~Akhoundi, ``Performance studies of
  underwater wireless optical communication systems with spatial diversity:
  {MIMO} scheme,'' \emph{IEEE Trans. Commun.}, vol.~65, no.~3, pp. 1176--1192,
  2017.

\bibitem{safari2008relay}
M.~Safari and M.~Uysal, ``Relay-assisted free-space optical communication,''
  \emph{IEEE Trans. Wireless Commun.}, vol.~7, no.~12, pp. 5441--5449, 2008.

\bibitem{chatzidiamantis2010generalized}
N.~D. Chatzidiamantis, G.~K. Karagiannidis, and M.~Uysal, ``Generalized
  maximum-likelihood sequence detection for photon-counting free space optical
  systems,'' \emph{IEEE Trans. Commun.}, vol.~58, no.~12, pp. 3381--3385, 2010.

\bibitem{lee2005diffuse}
Z.-P. Lee, M.~Darecki, K.~L. Carder, C.~O. Davis, D.~Stramski, and W.~J. Rhea,
  ``Diffuse attenuation coefficient of downwelling irradiance: An evaluation of
  remote sensing methods,'' \emph{Journal of Geophysical Research: Oceans},
  vol. 110, no.~C2, 2005.
\end{thebibliography}
\end{document}